\documentclass[a4paper,fleqn,usenatbib]{mnras}

\usepackage{newtxtext,newtxmath}

\usepackage[T1]{fontenc}
\usepackage{ae,aecompl}

\usepackage[dvipdfmx]{graphicx}	
\usepackage{amsmath}	
\usepackage{amssymb}	
\usepackage{indentfirst}	
\usepackage{color}
\usepackage{pict2e}

\newcommand{\mr}{\mathrm} 

\newcommand{\mbh}{M_\mr{BH}} 
\newcommand{\ninf}{n_\infty} 
\newcommand{\tinf}{T_\infty} 
\newcommand{\cinf}{c_\mr{s,\infty}} 

\newcommand{\mb}{\dot{M}_\mr{B}} 
\newcommand{\rb}{R_\mr{B}} 

\newcommand{\rj}{R_\mr{J}} 

\newcommand{\ledd}{L_\mr{E}} 
\newcommand{\medd}{\dot{M}_\mr{E}} 
\newcommand{\ledduv}{L_\mr{E,UV}} 
\newcommand{\leddir}{L_\mr{E,IR}} 
\newcommand{\meddir}{\dot{M}_\mr{E,IR}} 
\newcommand{\rhii}{R_\mr{HII}} 
\newcommand{\thii}{T_\mr{HII}} 
\newcommand{\nhii}{n_\mr{HII}} 

\newcommand{\teq}{T_\mr{eq}} 
\newcommand{\gsfr}{G_\mr{SFR}} 

\newcommand{\zcrit}{Z_\mr{E,crit}} 
\newcommand{\zdcrit}{Z_\mr{a,crit}} 

\newcommand{\mvir}{M_\mr{vir}} 
\newcommand{\rvir}{R_\mr{vir}} 
\newcommand{\mg}{M_\mr{g}} 
\newcommand{\rd}{R_\mr{d}} 
\newcommand{\sgmd}{\Sigma_\mr{d}} 
\newcommand{\zd}{z_\mr{d}} 
\newcommand{\nd}{n_\mr{d}} 

\newcommand{\mvten}{\left(\frac{\mvir}{10^{10} \ \mr{M}_\odot} \right)} 
\newcommand{\zten}{\left(\frac{z}{10} \right)} 
\newcommand{\tfour}{\left(\frac{T}{10^4 \ \mr{K}} \right)} 

\newcommand{\rsb}{R_\mr{sb}} 

\title[Dusty gas accretion onto seed BHs]
{Super-Eddington accretion of dusty gas onto seed black holes: 
metallicity-dependent efficiency of mass growth}

\author[Toyouchi et al.]{
Daisuke~Toyouchi$^1$, 
Takashi~Hosokawa$^1$, 
Kazuyuki~Sugimura$^2$, 
\newauthor
Riouhei~Nakatani$^3$, 
Rolf~Kuiper$^4$ \\
\\
$^{1}$Theoretical Astrophysics Group, Department of Physics, Kyoto University, Sakyo-ku, Kyoto 606-8502, Japan \\
$^{2}$Astronomical Institute, Tohoku University, Aoba-ku, Sendai 980-8578, Japan \\
$^{3}$Department of Physics, School of Science, The University of Tokyo, 7-3-1 Hongo, Bunkyo, Tokyo 113-0033, Japan \\
$^{4}$Institute of Astronomy and Astrophysics, University of T\"ubingen, Auf der Morgenstelle 10, D-72076 T\"ubingen, Germany
}

\date{Accepted XXX. Received YYY; in original form ZZZ}

\pubyear{2018}

\begin{document}
\label{firstpage}
\pagerange{\pageref{firstpage}--\pageref{lastpage}}
\maketitle

\begin{abstract}
The super-Eddington accretion onto intermediate seed BHs is a potential formation mode of supermassive black holes exceeding $10^9~M_\odot$ in the early universe. We here investigate how such rapid accretion may occur with finite amounts of heavy elements contained in the gas and dust. In our 1D radiation-hydrodynamics simulations, the radiative transfer is solved for both the direct UV lights emitted by an accretion disk and the diffuse IR lights thermally emitted by dust grains. Our results show that the radiative force by the IR lights causes a strong feedback to regulate the mass accretion. The resulting mean accretion rate is lower with the higher metallicity, and there is the critical metallicity $Z \sim 10^{-2} \ Z_\odot$, above which the super-Eddington accretion is prevented by the radiation pressure of the IR lights. With this taken into account, we examine if the dusty super-Eddington accretion occurs in young galaxies using a simple model. We show that a sufficient number of galaxies at $z \gtrsim 10$ can be such potential sites if BHs accrete the cold dense gas with $T \sim 10^2$~K, approximately the thermal equilibrium value at $Z = 10^{-2} \ Z_\odot$. We argue that the efficiency of the BH growth via the rapid accretion depends on the metallicity, and that the metallicity slightly lower than $10^{-2}~Z_\odot$ provides a chance for the most efficient growth.
\end{abstract}

\begin{keywords}
quasars: supermassive black holes -- radiation: dynamics
\end{keywords}

\section{INTRODUCTION} \label{sec:intro}

Unraveling the formation mechanism of super-massive black holes (SMBHs) is one of the big challenges in modern astrophysics. The recent discoveries of SMBHs with $\mbh \sim 10^9 \ M_\odot$ at high-redshift ($z \gtrsim 6$) provides a strict constraint for their formation timescale \citep[e.g.,][]{Fan01, Willott10, Mortlock11, Venemans13, Wu15, Banados18}. Accordingly, various seed BH scenarios have been suggested \citep[see, e.g.,][for a review]{Volonteri12, Haiman13}, including Pop III remnant BHs with mass of $\mbh \lesssim 10^3 \ M_\odot$ \citep[e.g.,][]{Yoshida08, Hosokawa11, Hosokawa16, Susa14, Hirano15, Stacy16}, massive BHs with $\mbh \sim 10^5 \ M_\odot$ formed via the direct collapse of supermassive stars \citep[e.g.,][]{Omukai01, Bromm03, Hosokawa12, Sugimura14, Sugimura16, Inayoshi14, Umeda16, Chon16, Chon18}, and intermediate mass BHs with $\mbh \sim 10^3 \ M_\odot$ formed as a consequence of stellar mergers in dense clusters \citep[e.q.,][]{Omukai08, Devecchi09, Katz15, Tagawa15, Yajima16, Sakurai17}. In any of these scenarios, seed BHs need to subsequently grow via super or almost Eddington accretion to become the observed SMBHs at $z \gtrsim 6$. However, whether such an efficient accretion can be realized in the universe has been actively under debate. 


Realistic gas accretion processes onto BHs have been investigated in many previous studies utilizing radiation hydrodynamics (RHD) simulations, solving the gas dynamics over the scale of the Bondi radius \citep[e.g.,][]{Milo09a, Milo09b, Park11, Park12, Park17}. 
These previous studies have shown that the gas accretion rate is generally suppressed under the Eddington rate by the thermal pressure enhanced due to the radiation feedback from the BH accretion disk. 
However, Inayoshi, Haiman and Ostriker \citep[2016, hereafter][]{Inayoshi16} show that, in sufficiently dense environments (e.g., $n_\mr{H} \sim 10^6 \ \mr{cm}^{-3}$ for $\mbh \sim 10^3 \ M_\odot$), gas accretion can proceed just like the Bondi flow, since the ram pressure of accreting gas overcomes the thermal and radiation pressure excess within the ionized region \citep[see also][]{Sakurai16}. 
In our paper, we call such a very efficient and stable mass accretion as {\it quasi-steady accretion}. \cite{Inayoshi16} suggests that such a quasi-steady accretion would occur in the early universe, and therefore they might contribute to the formation of SMBHs at $z \gtrsim 6$.


Although the scenario proposed by \cite{Inayoshi16} is intriguing, it still has limitations owing to assumptions made for simplicity. One of them is the isotropic radiative feedback from the central BH, while in reality the high-energy photons emitted from the inner part of accretion disk are preferentially attenuated along rays passing through the disk outer part \citep[e.g.,][]{Proga00, Proga04, Nomura16}. This effect has been examined in recent 2D-RHD simulations though anisotropic radiation fields are arbitrarily assumed \citep[][]{Sugimura17, Takeo18}. They suggest that the efficient accretion is even more easily realized than in the isotropic feedback cases, because the Bondi-like flow appears through horizontal layers shielded against the radiative feedback.
However, more recent work by \cite{Sugimura18} shows that only small amount of the angular momentum, with which the central accretion disk appears, substantially reduces the accretion rate. It is still uncertain how the above opposite effects compete with each other. 


Another simplification adopted in most of previous studies is only considering the primordial gas composition. In contrast, recent observations report various signatures of heavy elements in distant galaxies. For instance, \cite{Inoue16} present that a star-forming galaxy at $z \simeq 7.2$ already shows [O/H] $\sim -1$.  It has been also reported that bright quasar host galaxies at $z \gtrsim 6$ contain large amount of dust grains \citep[e.g.,][]{Venemans12, Venemans17}. Recent numerical simulations also predict that the metallicities at the galactic centers exceed $0.01 \ Z_\odot$ early on at $z \sim 10$ \citep[e.g.,][]{Ricotti05, Wise12, Ricotti16}. It is thus likely that SMBHs have grown via accreting such moderately metal-enriched gas at $z \gtrsim 6$. 
\cite{Yajima17} study the dynamics of the dusty accretion flow onto BHs with a suite of 1D RHD simulations, and show that the radiative force on dust grains regulates the mass accretion. However, they consider only cases with less dense environments, where the quasi-steady accretion does not occur.


In our current work, 
based on 1D-RHD simulations, we study effects of finite metallicities particularly for cases where the rapid BH mass growth via super-Eddington accretion is expected according to \cite{Inayoshi16}.
To represent the realistic dynamical structures of the dusty gas accretion, 
our RHD simulations calculate the radiative transfer and resulting radiative forces consistently for both the UV photons from the central accretion disk and the IR photons from thermal emissions of dust grains. 
Note that effects of the IR dust thermal emission is not included in \cite{Yajima17}.
We show that unlike the primordial case, the quasi-steady accretion does not necessarily lead to super-Eddington mass growth of BHs, since the radiative force via IR photons substantially modifies the gas accretion structures. 


The paper is organized as follows. In Section \ref{sec:method}, we describe the numerical method and settings of our simulations. In Section \ref{sec:bondi flow}, prior to our numerical results, we analytically investigate necessary conditions for the quasi-steady accretion of dusty gas. In Section \ref{sec:result}, we present the main results of our numerical simulations. 
With these results taken into account, we further investigate whether young galaxies provide potential sites of super-Eddington accretion in Section \ref{sec:gal}. 
In Section \ref{sec:discussion}, we provide discussions regarding physical processes which are not incorporated in our current simulations. Finally, the summary and conclusion are given in Section \ref{sec:summary}.

\section{SIMULATION METHOD} \label{sec:method}

In order to follow the dusty gas dynamics under the radiative feedback from a central BH, we make use of a modified version of the publicly available multi-dimensional MHD code PLUTO \citep[version 4.1;][]{Mignone07}. As in our previous studies \citep[e.g.,][]{Sugimura17, Nakatani18}, we consistently solve the hydrodynamics and radiative transfer equations, together with a non-equilibrium chemistry network. Various heating and cooling rates, which depends on the chemical abundances, are also included as source terms in the energy equation. We briefly describe our simulation method in what follows. Some more details are also described in \cite{Sugimura17}, who consider similar situations but for the primordial gas composition.

\subsection{Basic Equations} \label{sec:basiceq}

We perform a suite of one-dimensional hydrodynamics simulations to follow the dusty gas dynamics assuming spherical symmetry. We ignore effects of the angular momentum and magnetic fields that may be present with the ambient gas for simplicity. The basic equations are described as follows:
\begin{eqnarray}
\frac{\partial \rho}{\partial t} + \frac{1}{r^2}\frac{\partial}{\partial r}(r^2 \rho v) = 0,  
\label{eq:mass_cons}
\end{eqnarray}
\begin{eqnarray}
\rho \left ( \frac{\partial v}{\partial t} + v \frac{\partial v}{\partial r} \right ) = - \frac{\partial p}{\partial r} - \frac{G \mbh \rho}{r^2} + f_\mr{rad},  
\label{eq:mom_cons}
\end{eqnarray}
\begin{eqnarray}
\rho \left ( \frac{\partial e}{\partial t} + v \frac{\partial e}{\partial r} \right ) = - \frac{p}{r^2}\frac{\partial r^2 v}{\partial r} + \rho ( \Gamma - \Lambda),  
\label{eq:ene_cons}
\end{eqnarray}
\begin{eqnarray}
\frac{\partial n_\mr{H} y_i}{\partial t} + \frac{1}{r^2}\frac{\partial}{\partial r}(r^2 n_\mr{H} y_i v) = n_\mr{H} R_i,  
\label{eq:chem_cons}
\end{eqnarray}
where $\rho$, $v$, $p$ and $e$ are gas density, velocity, pressure, and specific energy of gas, $M_{\rm BH}$ the mass of a central BH, $f_\mr{rad}$ the radiative force (see Sec.~\ref{sec:rad_trans}), $\Gamma$ and $\Lambda$ the specific heating and cooling rates (see Sec.~\ref{sec:chem_therm}), $y_i$ and $R_i$ the ratio of the number density of $i$-th species to that of hydrogen nuclei and the corresponding chemical reaction rate. We here ignore the self-gravity of the gas for simplicity, and discuss its potential effects in detail in Section \ref{sec:multi}. 


We assume that dust grains are contained in the gas with the dust to gas mass ratio which is always constant in a calculation. This is equivalent to assuming a complete dynamical coupling between gas and dust. 
%
%
We do not include dust destruction processes such as the sputtering and sublimation, which are evaluated to be almost negligible in thermal states observed in our simulations. The dust sublimation potentially occurs in a very vicinity of the BH, which is currently outside of our computational domain (also see Section~\ref{sec:inner}). 
%
%

\subsection{Chemical and Thermal Processes} \label{sec:chem_therm}

In this study, we consider the eight chemical species: HI, HII, HeI, HeII, HeIII, CII, OI, and $e^-$.
We here ignore molecules, such as H$_2$ and CO, in the ambient medium. This is approximately valid for our cases with $Z \leq 10^{-1} \ Z_\odot$ (see Table 1), where the timescale of the molecule formation is generally much longer than the gas dynamical timescale  \citep{Krumholz12}.
We assume that the dust abundance is proportional to the gas metallicity $Z$, i.e., the dust to gas mass ratio is set as $0.01\times Z/Z_\odot$. The CII and OI abundances are also assumed to be constant at $y_\mr{CII} = 0.927 \times 10^{-4} Z/Z_\odot$ and $y_\mr{OI} = 3.568 \times 10^{-4} Z/Z_\odot$. Further ionization and recombination of CII and OI are not considered in our simulations.

We use almost the same chemistry network as in \cite{Sugimura17}, which includes photoionization and collisional ionization of HI, HeI and HeII, and recombination of HII, HeII, HeIII. Accordingly, the thermal processes considered are photoionization heating of HI, HeI and HeII, photoelectric heating of dust, recombination cooling of HII, HeII and HeIII, collisional ionization cooling of HI, HeI and HeII, fine-structure line cooling via CII and OI emission, free-free cooling of HI, HeI and HeII, and dust-gas collisional cooling. 
As far as we consider low metallicity environments with $Z < 10^{-1} \ Z_\odot$, other heating and cooling processes via heavy elements, i.e., meta-stable line cooling and photoionization heating, hardly affect our results \citep[e.g.,][]{Sutherland93, Dere09, Milo09b}.
Complete lists of our adapted chemical and thermal processes are presented in \cite{Sugimura17}, and also in \cite{Nakatani18} for the cooling processes via heavy elements.

\subsection{Subgrid model for the irradiation by BH} \label{sec:subgrid}

The computational domain of our simulations radially extends across the Bondi radius,
\begin{eqnarray}
\rb &=& \frac{G \mbh}{c^2_\mr{s,\infty}} \nonumber \\
&=& 1.4 \times 10^4~\mr{AU} \left ( \frac{\mbh}{10^3 \ \mr{M_\odot}} \right ) \left ( \frac{\tinf}{10^4 \ \mr{K}} \right )^{-1} \ , \nonumber \\
\label{eq:rbondi}
\end{eqnarray}
where we assume isothermal primordial gas, corresponding to the polytropic index $\gamma = 1$, the mean molecular weight $\mu = 1.3$, and the sound speed $\cinf = \sqrt{k_\mr{B} \tinf / (\mu m_\mr{p})} = 8.1 (\tinf/10^4 \ \mr{K})^{1/2} \ \mr{km \ s^{-1}}$. The inner and outer boundaries are located at $R_\mr{in} = 1/30\rb$ and $R_\mr{out} = 5\rb$, where the reference Bondi radius is estimated for $\tinf = 10^4 \ \mr{K}$. 
We separate the radial space of our computational domain into 258 logarithmic grid cells. 
We have confirmed that further increasing the grid numbers does not affect our results. 
We separately discuss effects of varying the position of the inner boundary in Section \ref{sec:inner}.

The inside of $R_\mr{in}$ is a sink region, where we suppose that a circum-BH accretion disk forms. Although we assume no angular momentum in the accreting gas, the situations we consider represent cases where the size of the accretion disk is much smaller than the central sink region (see Section \ref{sec:multi} for further discussions). A mass accretion rate onto the accretion disk $\dot{M}$ is evaluated with the inward mass flux measured at $R_\mr{in}$. Then photons emitted from the central BH are injected from the inner boundary depending on $\dot{M}$ at each time step. To transform $\dot{M}$ to the luminosity $L$, we adopt the fitting formula given by \cite{Watarai00},
\begin{eqnarray}
L =  
\begin{cases}
2 \ \ledd \ \left [ 1 + \mr{ln} \left ( \frac{\dot{m}}{2} \right ) \right ] & (\dot{m} > 2) \\
\ledd \ \dot{m} & (\mr{otherwise})
\end{cases}
\label{eq:MtoL}
\end{eqnarray}
where $\ledd$ is the Eddington luminosity 
\begin{eqnarray}
\ledd 
= 3.3 \times 10^7~L_\odot \left ( \frac{\mbh}{10^3 \ \mr{M_\odot}} \right ) \ ,
\label{eq:ledd}
\end{eqnarray}
and $\dot{m}$ is defined as $\dot{M}/\medd$ with the Eddington mass accretion rate
\begin{eqnarray}
\medd &=& \frac{\ledd}{\eta c^2} \nonumber \\
&=& 2.2 \times 10^{-5}~M_\odot \ {\rm yr}^{-1} 
\left ( \frac{\mbh}{10^3 \ \mr{M_\odot}} \right )
\left( \frac{\eta}{0.1} \right)^{-1} \ .
\label{eq:medd}
\end{eqnarray}
Unless noticed, we set $\eta=0.1$ throughout this paper. Finally, note that we do not allow mass ejection (or outflow) from the central sink region for simplicity.

\subsection{Radiative Transfer} \label{sec:rad_trans}

We solve the frequency-dependent radiative transfer of far ultraviolet (FUV; $6 \ \mr{eV} < \nu < 13.6 \ \mr{eV}$) and extreme ultraviolet (EUV; $\nu > 13.6 \ \mr{eV}$) photons. 
As in previous studies \citep{Milo09b, Park11, Park12, Sugimura17}, we inject photons with a power-law spectrum $L_\nu \propto \nu^{-\alpha} (\alpha = 1.5)$ between $\nu_\mr{min} = 6~{\rm eV}$ and  $\nu_\mr{max}= 1~{\rm keV}$ from the sink region to the innermost cell. 
The spectrum is normalized so that its integration over $\nu_\mr{min} < \nu < \nu_\mr{max}$ is equal to the total luminosity $L$ given by equation (\ref{eq:MtoL}). We use 128 logarithmically spaced frequency bins for calculations presented below. 
%
%
Note here that the total energy ratio of FUV to EUV photons depends on the spectrum index $\alpha$. Since recent AGN observations show that the value of $\alpha$ generally distributes from 1.5 to 2.5 \citep[e.g.,][]{Risaliti09, Ueda14, Kawamuro16, Trakhtenbrot17}, we have performed several test simulations with the same settings as described below but for $\alpha = 2.5$. We have confirmed that the uncertainty on $\alpha$ does not affect our conclusions. 
%
%


We consider the radial propagation of FUV and EUV photons coming from a central accretion disk.
The direct EUV photons are consumed to photoionize HI, HeI, and HeII, for which we use frequency-dependent cross sections given by \cite{Osterblock89} and \cite{Yan98}. Additionally, we take into account the attenuation of FUV and EUV photons by dust absorption with opacity table of \cite{Weingartner01}. Note that we do not count potential consumption of FUV photons via photodissociation because we assume no molecules present in the ambient medium. The radiative force via Thomson scattering, photoionization, and dust absorption are consistently evaluated with the radial radiative transfer described above.


In addition to the FUV and EUV components described above, we further consider the transfer of diffuse IR radiation produced by thermal dust emission. The governing equation is written as
\begin{eqnarray}
\frac{\partial E_\mr{rad}}{\partial t} + \frac{1}{r^2}\frac{\partial}{\partial r}(r^2 F_\mr{IR}) = \kappa_\mr{P} \rho c (a_\mr{R} T_\mr{d}^4 - E_\mr{rad}),  
\label{eq:erad_cons}
\end{eqnarray}
where $E_\mr{rad}$ denotes energy density of IR radiation, $T_\mr{d}$ dust temperature, $F_\mr{IR}$ the IR radiative flux, $\kappa_\mr{P}$ the Planck mean opacity, $a_\mr{R}$ the radiation constant, and $c$ the speed of light. In our model, $F_\mr{IR}$ is described with the flux-limited diffusion (FLD) approximation as
\begin{eqnarray}
F_\mr{IR} = - \frac{\lambda c}{\kappa_\mr{R} \rho} \frac{1}{r^2}\frac{\partial}{\partial r} (r^2 E_\mr{rad}),
\label{eq:fld}
\end{eqnarray}
where $\kappa_\mr{R}$ is the Rosseland mean opacity and $\lambda$ is the flux-limiter, which we take from \cite{Levermore81}. 
%
%
Note that FLD approximation is applied only for the component of the diffuse IR radiation. The other direct component coming from the accreting BH is separately solved using the ray-tracing method. It has been shown that such a hybrid scheme provides the much more accurate approximation than using either of them. Although the FLD is not strictly accurate especially for cases where the medium is moderately optically thick for the infrared light with $Z = 10^{-3} \ Z_\odot$ (see Section \ref{sec:initial}), we believe that such an effect is minor. In our current 1D simulations assuming the spherical symmetry, the radiative transport is quite simple and the flux is just approximated with $F_\mr{IR} \sim L/4\pi r^2$.
%
%

Equation (\ref{eq:erad_cons}) is numerically solved by combining with the equation to determine the dust temperature
\begin{eqnarray}
c_\mr{v} \frac{\partial T_\mr{d}}{\partial t} = - \kappa_\mr{P} c (a_\mr{R} T_\mr{d}^4 - E_\mr{rad}) + S ,  
\label{eq:td_cons}
\end{eqnarray}
where S is the source term representing the heating of dust grains via absorptions of FUV and EUV photons, and $c_\mr{v}$ is the specific heat capacity given by $c_\mr{v} = k_\mr{B}/(\gamma-1)\mu m_\mr{p}$. To solve equations (\ref{eq:erad_cons}) and (\ref{eq:td_cons}), we make use of the radiative transport module developed in \cite{Kuiper18}, which has been primarily used to study present-day high-mass star formation \citep[e.g.,][]{Kuiper10, Kuiper11, Kuiper12}. The obtained dust temperature $T_{\rm d}$ is used to provide the dust-gas collisional cooling rate in the gas energy equation (\ref{eq:ene_cons}).


We include the acceleration by the radiation pressure force owing to the diffuse IR component in the gas momentum equation (\ref{eq:mom_cons}),
\begin{eqnarray}
f_\mr{rad,IR} =  \frac{\rho \kappa_\mr{R} F_\mr{IR}}{c}.
\label{eq:f_ir}
\end{eqnarray}
It is worth noting here that because the dust opacity for IR photons is much smaller than that for UV photons, IR radiative force is basically minor in an HII region, where direct UV photons come to produce the much stronger radiative force. Unlike UV photons, however, IR photons can penetrate deep into the neutral medium. The IR radiative force thus can potentially modify the accretion flow structure outside of the HII region. We will show that this effect requires another condition for the super-Eddington accretion onto BHs embedded in the dusty gas (see Sec.~\ref{sec:rad_ir} below).


\subsection{Cases Considered and Initial Conditions} \label{sec:initial}

We suppose gas accretion onto a BH with mass $\mbh$, residing in homogeneous gas with number density $\ninf$, temperature $\tinf$, and metallicity $Z$. We consider several cases with different parameter sets as summarized in Table \ref{table:model}. For the cases of M4UV, M5UV and M5UVIR, we perform the simulations starting with the same initial conditions but for different metallicities $Z/Z_\odot = 10^{-3}, \ 10^{-2}$, and $10^{-1}$, to investigate metallicity dependencies.


The primordial gas composition has been assumed to investigate the rapid accretion onto seed BHs mostly in the previous studies \citep[e.g.,][]{Milo09b, Park11, Park12, Inayoshi16}. In these studies, the ambient temperature $\tinf$ has been fixed at $\sim 10^4$~K, below which no efficient cooling processes operate if molecules are not present. In our cases, however, the ambient temperature may be much lower than $10^4$~K and metallicity-dependent owing to additional heating and cooling processes. We nonetheless assume $T_\infty \sim 10^4$~K for many cases (e.g., M5UV and M5UVIR) even with non-zero metallicities for the following reasons. Firstly, we aim to isolate the effects of UV and IR radiative forces. We thus separately examine potential impacts of varying the ambient temperature depending on metallicities (model M3UVIR\_C, see Section \ref{sec:mcool}). Secondly, we take account of the observational fact that the ISM is filled with turbulent motions in various environments. The turbulence is in general supersonic over $\gtrsim$~pc spatial scale, for which we study the accretion flow onto a BH. Although the turbulence can not be fully modeled in our 1D calculations, we adopt $T_\infty \sim 10^4$~K to represent the ``effective temperature'' of such non-thermal turbulent medium. We further discuss potential roles of the turbulence in the accreting gas in Section \ref{sec:multi}.


\begin{table*}
\begin{center}
\caption{Model parameters} \label{table:model}
\begin{tabular}{ccccccc} \hline \hline
Model      & $\mbh \ \rm [M_\odot]$ & $\ninf \ \rm [cm^{-3}]$ & $\tinf \ \rm [K]$ & $Z \ [\mr{Z_\odot}]$                       & Radiative force  & Metal cooling \\ \hline
M4UV       & $10^4$                          & $10^5$                       & $10^4$             & $10^{-3}, \ 10^{-2}, \ 10^{-1} $       & UV                     & NO  \\ 
M5UV       & $10^5$                          & $10^5$                       & $10^4$             & $10^{-3}, \ 10^{-2}, \ 10^{-1} $       & UV                     & NO  \\ 
M5UVIR         & $10^5$                          & $10^5$                       & $10^4$             & $10^{-3}, \ 10^{-2}, \ 10^{-1} $       & UV\&IR              & NO  \\
M3UVIR     & $10^3$                          & $10^5$                       & $10^4$            & $10^{-2} $                                      & UV\&IR              & NO  \\ 
M3UVIR\_C       & $10^3$                          & $10^5$                       & $\teq$              & $10^{-2} $                                      & UV\&IR              & YES \\ \hline
\end{tabular}
\end{center}
\end{table*}

\section{CONDITION FOR QUASI-STEADY accretion} \label{sec:bondi flow}

\cite{Inayoshi16} show that, with the primordial gas composition, the quasi-steady super-Eddington accretion onto a BH is realized when the Bondi radius is larger than the typical size of an HII bubble created around the BH. Before addressing our numerical results, we here consider how this condition may be modified if the accreting gas contains dust grains. Particularly in this section, we focus on the role of UV radiative force studied by \cite{Yajima17}.


Consider an HII bubble in a pressure equilibrium with the outer neutral medium with uniform-density $\ninf$ and temperature $\tinf$. In this case, the mean gas number density within the HII bubble is written as $\nhii = (\tinf / 2 \thii) \ \ninf$. The so-called Str\"omgren radius provides the size of the HII bubble,  
\begin{eqnarray}
\rhii &=& \left ( \frac{3Q_\mr{ion}}{4 \pi \alpha_\mr{rec,B} n^2_\mr{HII}} \right )^{1/3} \nonumber \\
&\propto& L^{1/3} T^{1/3}_\mr{HII} n^{-2/3}_\mr{HII} \ ,
\label{eq:rhii}
\end{eqnarray}
where $\alpha_\mr{rec} (\propto \thii)$ is the case B hydrogen recombination coefficient, $Q_\mr{ion} (\propto L)$ the emissivity of ionizing photons. Owing to the dependence of $\rhii \propto L^{1/3}$, the maximum size of the HII bubble is limited by the maximum luminosity, e.g., Eddington luminosity. For the case with the dusty gas, the Eddington luminosity is determined by the radiative force not only with Thomson scattering, but also with dust absorption of UV photons \citep[e.g.,][]{Yajima17}. The dust opacity for UV photons is given by $\kappa_\mr{d,UV} = 2.8 \times 10^{2} \ (Z/Z_\odot) \ \mr{cm^2 \ g^{-1}}$ for grains with the typical particle size of $a_\mr{d} \sim 0.1 \ \mu$m, which is the case with the dust model assumed in our simulations. A modified expression of the Eddington luminosity is thus given by 
\begin{eqnarray}
\ledduv &=& \frac{4 \pi G \mbh c}{\kappa_\mr{T} + \kappa_\mr{d,UV}} \nonumber \\
&=& \left \{ 1 + 7.1 \times (Z/10^{-2} \ Z_\odot) \right \}^{-1} \ \ledd \ ,
\label{eq:ledduv}
\end{eqnarray}
which shows that $\ledduv$ decreases with increasing metallicity for $Z \gtrsim 10^{-3} \ Z_\odot$, where the radiative force is boosted by dust absorption of UV photons. As already noted in Section \ref{sec:rad_trans}, the IR radiative force on dust is almost negligible in comparison to the UV force here. Then the maximum size of HII bubble is obtained as
\begin{eqnarray}
R_\mr{HII,max} &=&  4.0 \times 10^5~\mr{AU} \left(1 + 7.1 \times \frac{Z}{10^{-2} \ Z_\odot} \right)^{-1/3}  \nonumber \\
& & \ \times \left ( \frac{\thii}{7 \times 10^4 \ \mr{K}} \right ) \left ( \frac{\tinf}{10^4 \ \mr{K}} \right )^{-2/3} \nonumber \\
& & \ \times \left ( \frac{\mbh}{10^3 \ M_\odot} \right )^{1/3} \left ( \frac{\ninf}{10^5 \ \mr{cm^{-3}}} \right )^{-2/3} \ . \nonumber \\
\label{eq:rhiimax}
\end{eqnarray}
Eqs. (\ref{eq:ledduv}) and (\ref{eq:rhiimax}) suggest that, for $Z \gtrsim 10^{-3} \ Z_\odot$, the maximum size of the HII bubble is smaller with higher metallicity, because the maximum luminosity is reduced owing to the UV radiative force exerted on dust grains. 

\begin{figure}
\begin{center}
\includegraphics[width=8cm,height=6cm]{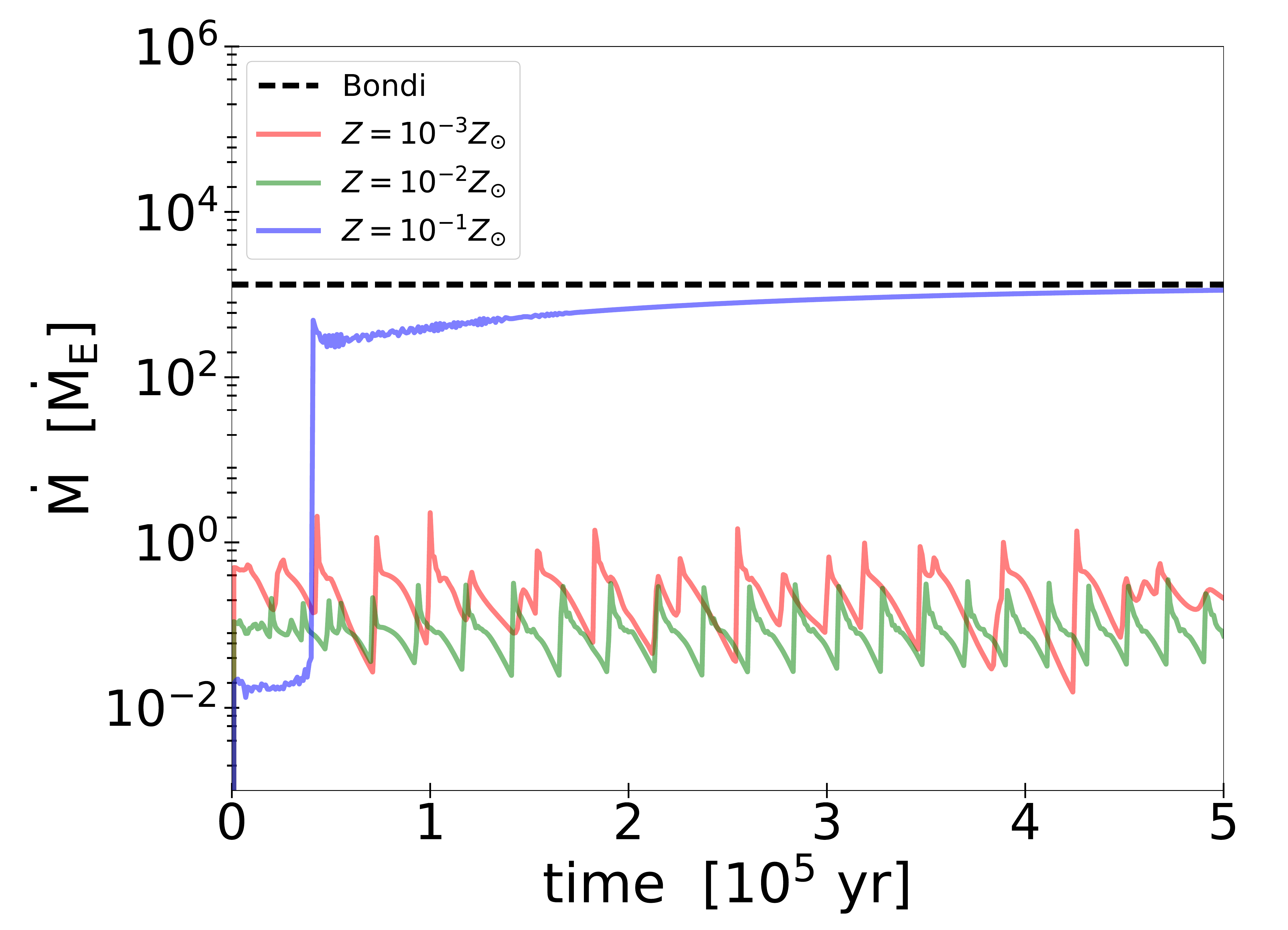}
\end{center}
\caption{Time evolution of accretion rates onto the BH for M4UV models.
The unit for the accretion rates is the Eddington rate $\dot{M}_{\rm E}$ defined by Eq. (\ref{eq:medd}).
The accretion rates averaged over $10^3$ yrs are presented in the figure.
The red, green, and blue lines correspond to the cases with different metallicities $Z/\mr{Z_\odot} = 10^{-3}, \ 10^{-2}, \ 10^{-1}$, respectively. The black horizontal dashed line represents the Bondi accretion rate given by equation (\ref{eq:mbondi}).}
\label{fig:M4UV}
\end{figure}

Here we note that, in the estimate of Eq. (\ref{eq:rhiimax}), consumptions of ionizing photons due to dust absorption is not considered. To check the validity of this treatment, we evaluate the critical metallicity, above which the optical depth with dust absorption across the HII bubble $\tau_\mr{d,UV} = \kappa_\mr{d,UV} \rho_\mr{HII} \rhii$ exceeds unity, 
\begin{eqnarray}
\zdcrit &\sim& 0.2~Z_\odot \left( \frac{\mbh}{10^3 \ M_\odot} \right)^{-1/2}  \nonumber \\
&\times& \left ( \frac{\ninf}{10^5 \ \mr{cm^{-3}}} \right )^{-1/2} 
\left ( \frac{\tinf}{10^4 \ \mr{K}} \right )^{-1/2} \ . \nonumber \\
\label{eq:zdcrit}
\end{eqnarray}
For $Z > \zdcrit$, the size of the HII bubble is expected to be determined mainly by dust absorption as
\begin{eqnarray}
R_\mr{HII} &\sim& (\kappa_\mr{d,UV} \rho_\mr{HII})^{-1} \nonumber \\
&\sim& 1.5 \times 10^6~\mr{AU} \left (  \frac{Z}{10^{-2} \ Z_\odot} \right )^{-1} \nonumber \\
&\times& \left ( \frac{\ninf}{10^5 \ \mr{cm^{-3}}} \right )^{-1} \left ( \frac{\tinf}{10^4 \ \mr{K}} \right )^{-1} \left ( \frac{\thii}{7 \times10^4 \ \mr{K}} \right ) . \nonumber \\
\label{eq:rhii_dust}
\end{eqnarray}
In summary, the condition for $\rb > \rhii$, where a quasi-steady accretion is expected to occur, can be described from Eqs. (\ref{eq:rbondi}), (\ref{eq:rhiimax}) and (\ref{eq:rhii_dust}) as follows:
\begin{eqnarray}
& & \left ( \frac{\mbh}{10^5 \ M_\odot} \right ) \left ( \frac{\ninf}{10^5 \ \mr{cm^{-3}}} \right ) \gtrsim \ \ \ \ \ \ \ \ \ \ \ \ \ \ \ \ \ \ \ \ \ \ \ \ \ \ \ \ \  \nonumber \\
& & \left ( \frac{\thii}{7 \times 10^4 \ \mr{K}} \right )^{3/2} \left ( \frac{\tinf}{10^4 \ \mr{K}} \right )^{1/2} \left(1 + 7.1 \times \frac{Z}{10^{-2} \ Z_\odot} \right)^{-1/2} \ , \nonumber \\
& & \ \ \ \ \ \ \ \ \ \ \ \ \ \ \ \ \ \ \ \ \ \ \ \ \ \ \ \ \ \ \ \ \  \ \ \ \ \ \ \ \ \ \ \ \ \ \ \ \ \ \ \ \ (\mr{for} \ Z < \zdcrit) \nonumber \\
\label{eq:condi1}
\end{eqnarray}
\begin{eqnarray}
& & \left ( \frac{\mbh}{10^5 \ M_\odot} \right ) \left ( \frac{\ninf}{10^5 \ \mr{cm^{-3}}} \right ) \gtrsim \left ( \frac{\thii}{7 \times 10^4 \ \mr{K}} \right ) \left(\frac{Z}{10^{-2} \ Z_\odot} \right)^{-1} \ . \nonumber \\
& & \ \ \ \ \ \ \ \ \ \ \ \ \ \ \ \ \ \ \ \ \ \ \ \ \ \ \ \ \ \ \ \ \  \ \ \ \ \ \ \ \ \ \ \ \ \ \ \ \ \ \ \ \ (\mr{for} \ Z > \zdcrit) \nonumber \\
\label{eq:condi2}
\end{eqnarray}
The condition of Eq. (\ref{eq:condi1}) for $Z < \zdcrit$ is an extension of that derived in \cite{Inayoshi16}\footnote{Eq. (\ref{eq:condi1}) in the case of $Z = 0$ is actually different from the corresponding condition, $(\mbh/10^4 \ M_\odot)(\ninf/10^5 \ \mr{cm^{-3}}) \gtrsim (\tinf/10^4 \ \mr{K})^{3/2}$, in \cite{Inayoshi16}. This is because they simply assume $\nhii = \ninf$ in estimation of $\rhii$.}, suggesting that the more massive BH and/or the denser and colder ambient medium are preferred to realize the quasi-steady super-Eddington accretion. Eq. (\ref{eq:condi2}) for $Z > \zdcrit$ generally shows the same trend as in Eq. (\ref{eq:condi1}) but for the absence of the $T_\infty$-dependence. Moreover, Eqs. (\ref{eq:condi1}) and (\ref{eq:condi2}) clearly show that the high metallicity is favored to cause the quasi-steady accretion, because HII bubble has the smaller size as found in Eqs. (\ref{eq:rhiimax}) and (\ref{eq:rhii_dust}). We confirm this metallicity dependence in numerical simulations for the cases of M4UV model. With the corresponding parameter sets $(\mbh, \ \ninf, \ \tinf) = (10^4 \ M_\odot, \ 10^5 \ \mr{cm^{-3}}, \ 10^4 \ \mr{K})$, Eqs. (\ref{eq:condi1}) and (\ref{eq:condi2}) predict that the quasi-steady accretion should occur for $Z \gtrsim 10^{-1} \ Z_\odot$. Figure \ref{fig:M4UV} presents the time evolution of gas accretion rates onto the BH in the unit of $\medd$ for the cases with $Z = 10^{-3}, \ 10^{-2}$, and $10^{-1}~Z_\odot$. For a reference, we also plot the Bondi accretion rate
\begin{eqnarray}
\mb &=& 1.7 \times 10^{-3} M_\odot \ \mr{yr}^{-1} \nonumber \\
& & \ \ \ \times
\left ( \frac{\ninf}{10^5 \ \mr{cm}^{-3}} \right ) \left ( \frac{\mbh}{10^3 \ \mr{M_\odot}} \right )^2 \left ( \frac{\tinf}{10^4 \ \mr{K}} \right )^{-3/2}. 
\label{eq:mbondi}
\end{eqnarray}
Figure \ref{fig:M4UV} shows that, for $Z = 10^{-3}$ and $10^{-2}~Z_\odot$, the mass accretion rates periodically oscillate throughout the runs, whereas for $Z = 10^{-1} \ Z_\odot$ it dramatically increases at $t \simeq 4 \times 10^4$ yr and eventually reach the quasi-steady accretion. These results are remarkably in good agreement with the predictions by Eqs. (\ref{eq:condi1}) and (\ref{eq:condi2}).

Another important finding in Figure \ref{fig:M4UV} is that although the radiative force is substantially boosted via dust absorption of UV photons, the mass accretion rate in the quasi-steady state is almost equal to the Bondi value, the same as in the cases with the primordial gas \citep[e.g.,][]{Inayoshi16, Sakurai16}. These facts indicate that the presence of dust grains in the gas promotes a transition to quasi-steady flow and consequently lead to super-Eddington accretion. However, recall that the IR radiative force is not included in M4UV models presented in Figure \ref{fig:M4UV}. In the next section we clarify how the IR radiative force changes the results.

\begin{figure}
\begin{center}
\includegraphics[width=8cm,height=10cm]{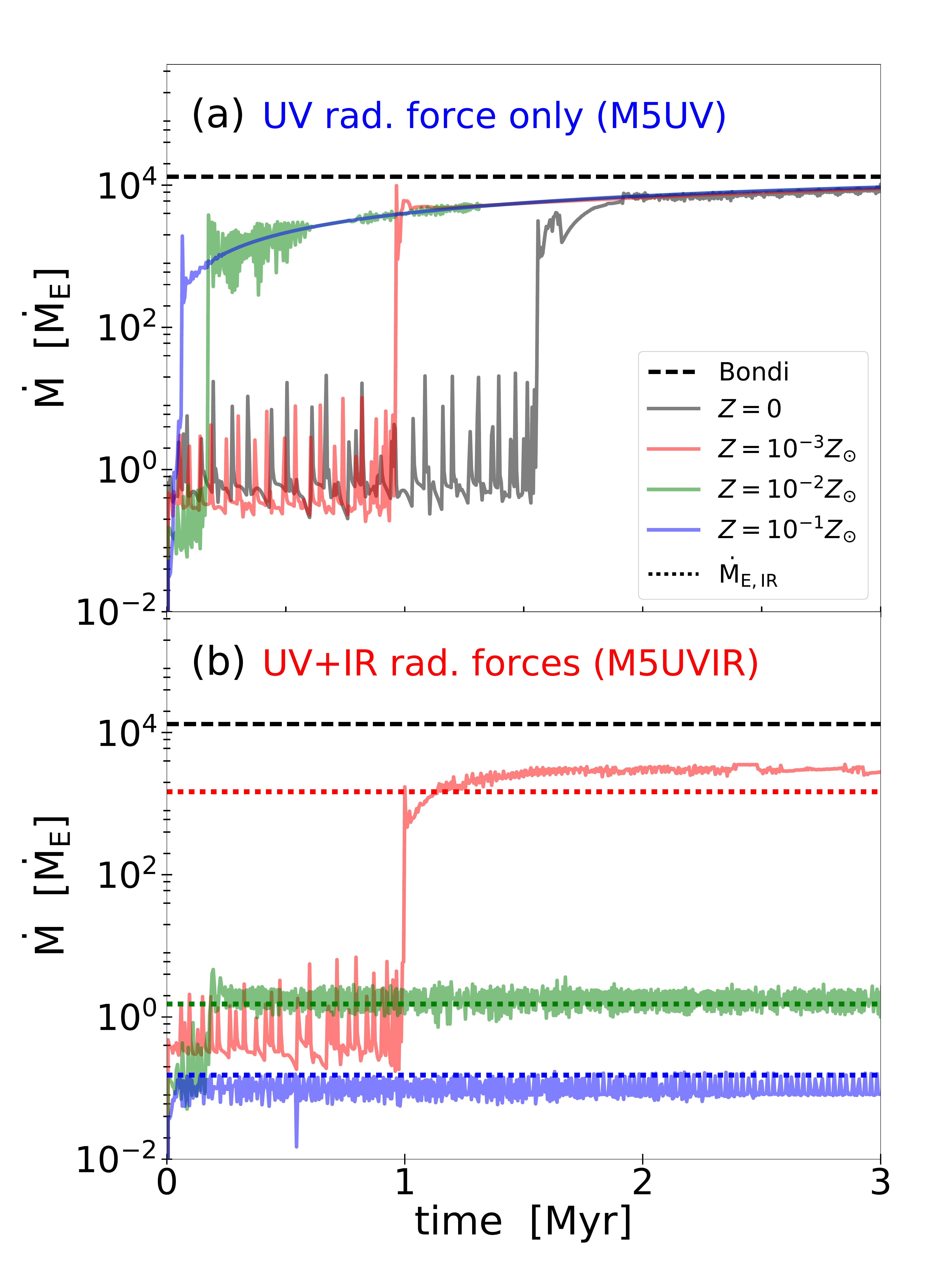}
\end{center}
\caption{
The same as Figure~\ref{fig:M4UV} but for M5UV models (panel a, only with the UV radiative force) and M5UVIR models (panel b, with both the UV and IR forces). The presented accretion rates are time-averaged for $10^4$ yrs. The horizontal dotted lines in the panel (b) represent the Eddington rates defined with the IR dust opacity $\dot{M}_{\rm E,IR}$ (Eq.~\ref{eq:meddir}). 
The different colors denote the different metallicities as in Figure~\ref{fig:M4UV}.
}
\label{fig:M5UVIR}
\end{figure}

\begin{figure}
\begin{center}
\includegraphics[width=8cm,height=10cm]{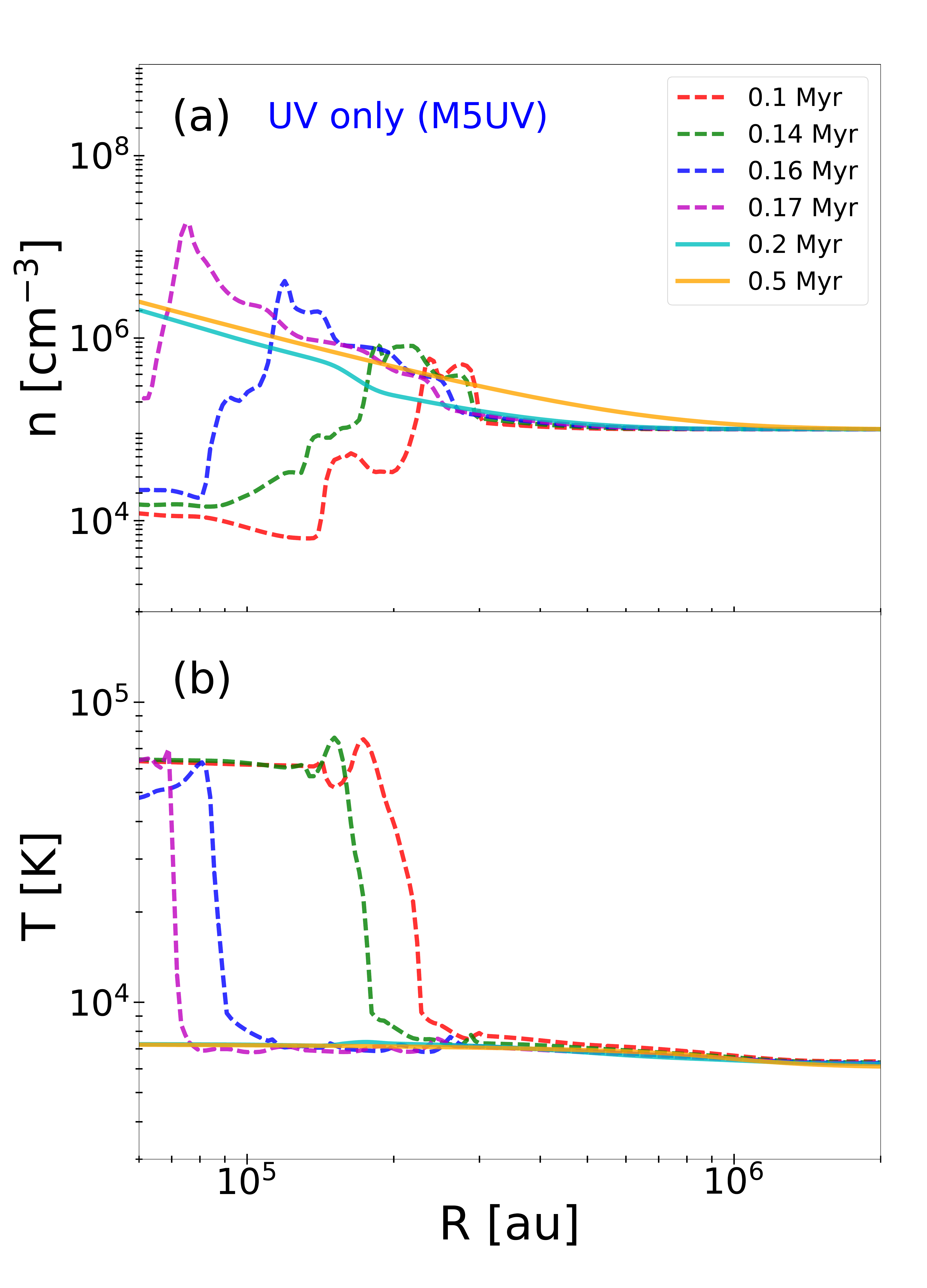}
\end{center}
\caption{Time evolution of the density (panel a) and temperature (panel b) structure of the accreting gas for M5UV model with $Z = 10^{-2} \ Z_\odot$.  The different colors represent the time sequence as shown at the upper right corner in panel (a).  The dashed and solid lines denotes the snapshots before and after the rapid increase of mass accretion rate, i.e., the transition into the quasi-steady accretion stage.}
\label{fig:den_woir}
\end{figure}

\begin{figure}
\begin{center}
\includegraphics[width=8cm,height=10cm]{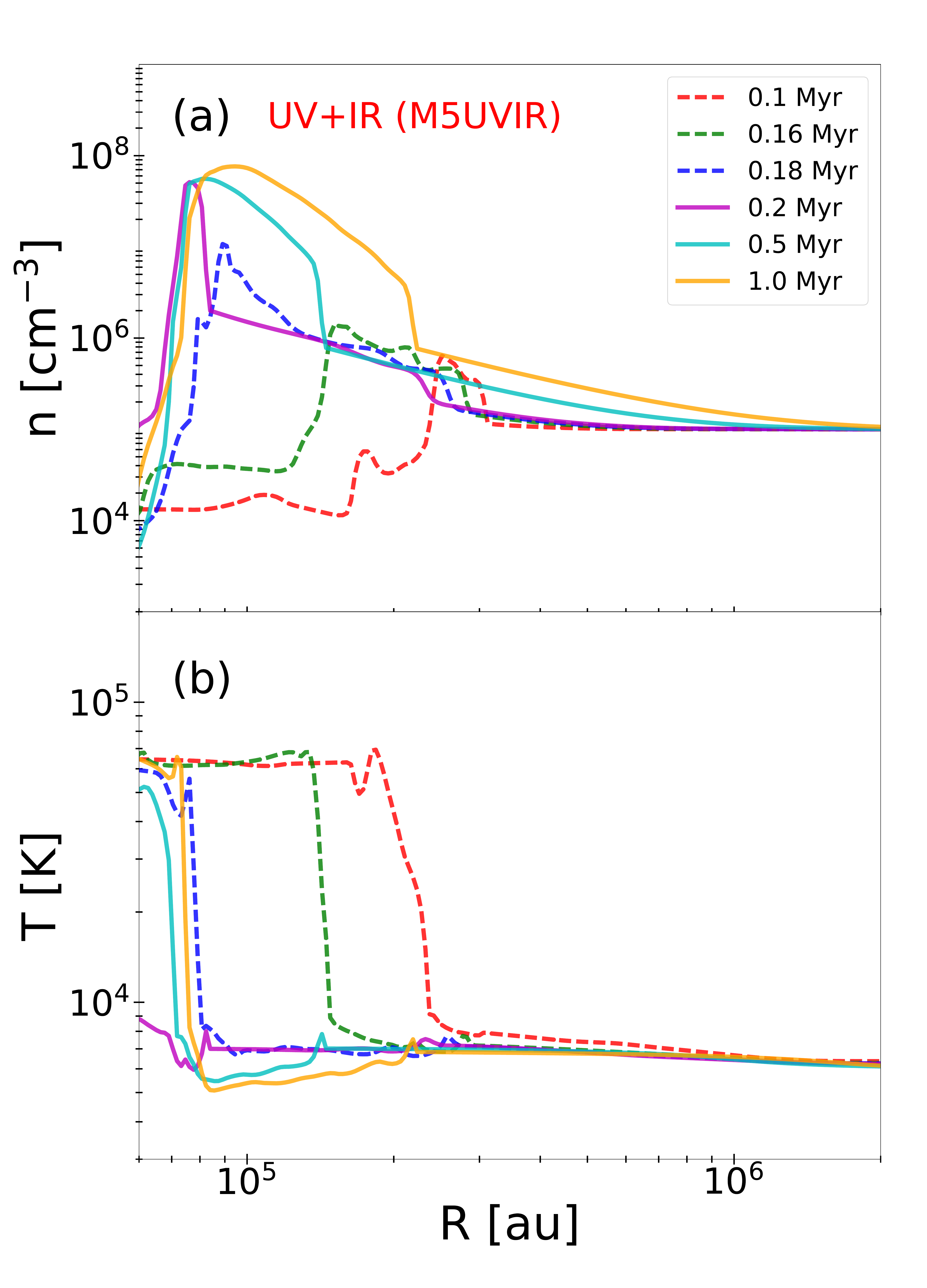}
\end{center}
\caption{The same as Figure \ref{fig:den_woir}, but for M5UVIR model with $Z = 10^{-2} \ Z_\odot$.}
\label{fig:den_wir}
\end{figure}

\begin{figure}
\begin{center}
\includegraphics[width=9cm,height=12cm]{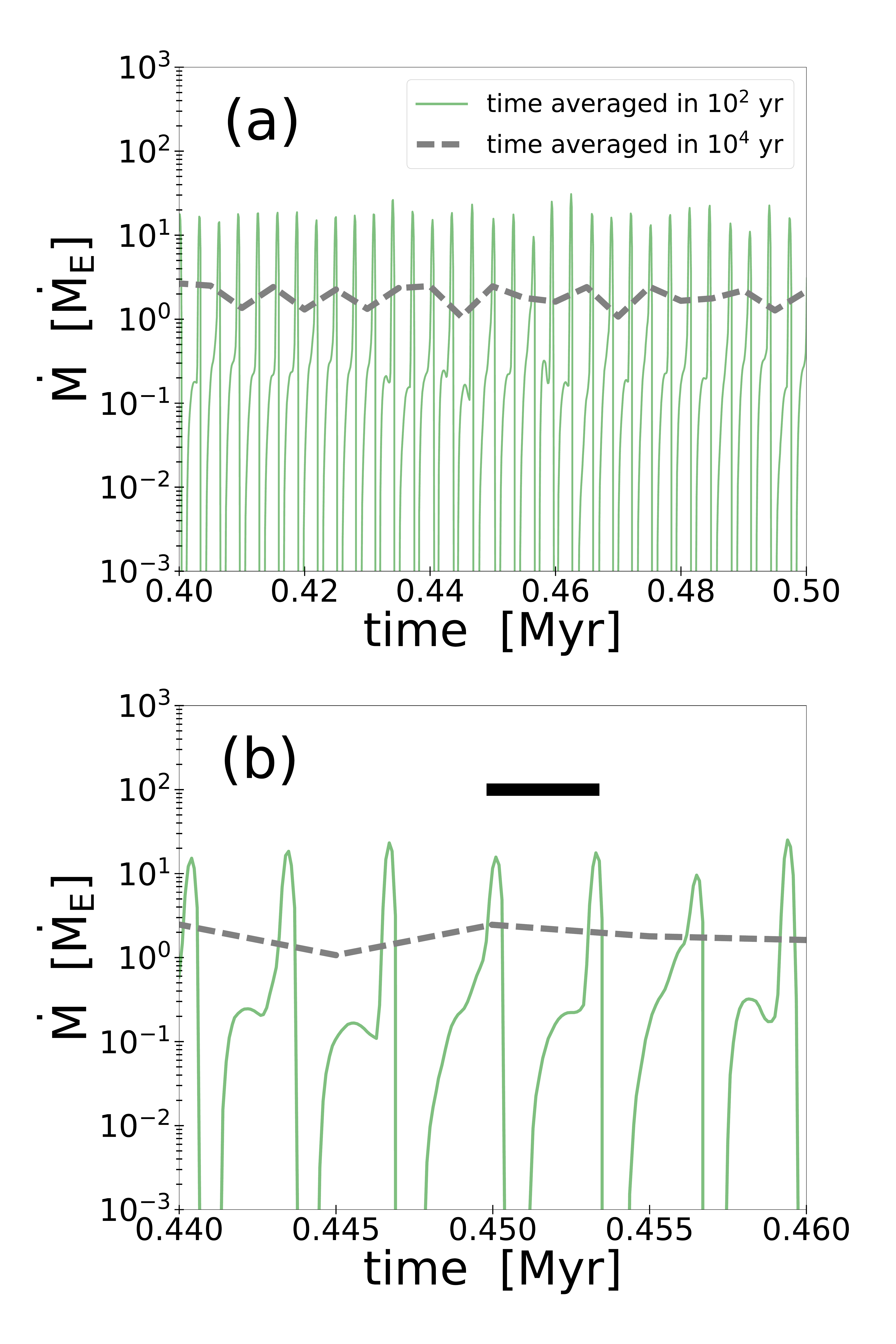}
\end{center}
\caption{Detailed behavior of the time variable accretion rates in the quasi-steady accretion stage in M5UVIR model with $Z = 10^{-2} \ Z_\odot$. Panels (a) and (b) show the time evolution of the accretion rates in short duration of $0.4 \leq t/\mr{Myr} \leq 0.5$ and $0.44 \leq t/\mr{Myr} \leq 0.46$. 
In the both panels, the thick gray and thin green lines correspond to the accretion rates averaged over $10^4$ and $10^2$ yrs, respectively. 
In panel (b) the horizontal thick black line segment denotes the period of $0.45 \lesssim t/\mr{Myr} \lesssim 0.453$, for which the detailed gas accretion structure is investigated in Figure \ref{fig:Force}. }
\label{fig:mdot}
\end{figure}

\section{RESULTS} \label{sec:result}

\subsection{Effects of IR Radiative Force} \label{sec:rad_ir}

As briefly noted in Section \ref{sec:rad_trans}, the IR radiative force becomes important for cases with $\rb > \rhii$, where the transition into the quasi-steady accretion is expected.
To investigate such an effect, we here thoroughly compare models M5UV and M5UVIR. 
The parameter set common for these models is $(\mbh, \ \ninf, \ \tinf) = (10^5 \ M_\odot, \ 10^5 \ \mr{cm^{-3}}, \ 10^4 \ \mr{K})$, which satisfies the condition $\rb > \rhii$ with any metallicities of $Z \geq 0$ (see Eqs. \ref{eq:condi1} and \ref{eq:condi2}). In model M5UVIR, the only difference from M5UV is the addition of the IR radiative force. 


In Figure \ref{fig:M5UVIR}, we present the time evolution of the mass accretion rates for models M5UV (panel a) and M5UVIR (panel b) in comparison.
Figure \ref{fig:M5UVIR} (a) shows that, for all of the M5UV models, the accretion rates rapidly increase at some points and gradually converge to the Bondi value.
The time variation of the accretion rate qualitatively changes across that, from the oscillatory stage to the quasi-steady stage for all the metallicities.
Such an evolution is consistent with the predictions given by Eqs. (\ref{eq:condi1}) and (\ref{eq:condi2}). Moreover, Figure \ref{fig:M5UVIR} (a) also shows that the transition to the quasi-steady phase occurs earlier with the higher metallicity. 
This is because the smaller HII bubble with the higher metallicity, as already noted by Eqs. (\ref{eq:rhiimax}) and (\ref{eq:rhii_dust}), leads to quicker gas accumulations onto the HII bubble surface. 
Figure \ref{fig:M5UVIR} (b) presents how the above results change with including the IR radiative force. 
Although M5UVIR models also show the rapid increases of accretion rates similarly to M5UV models, the mass accretion rates do not go to the Bondi rate.
The resulting values are much lower than the Bondi value, and lower with higher metallicity \footnote{The gas accretion after the rapid collapse of HII bubble in M5UVIR model is also called the quasi-steady accretion phase in this paper, regardless of the converged mass accretion rates.}. 


We also compare the time evolution of density and thermal structure of accreting gas for M5UV and M5UVIR models. Figures \ref{fig:den_woir} and \ref{fig:den_wir} present such a comparison with the particular cases with $Z = 10^{-2} \ Z_\odot$. Figure \ref{fig:den_woir} shows the evolution before and after the transition to the quasi-steady phase for M5UV model.
We see that the HII bubble, where the temperature is $5-7 \times 10^4$~K, gradually shrinks for $t \lesssim 0.18 \ \rm Myr$, i.e., before the transition occurs (dashed lines). 
The bubble is completely quenched and disappears in the end. The small HII region should be present trapped within the accretion flow, but it is not spatially resolved but masked by the central sink region we employ. The density profiles after the transition approximately follow the power-law distributions (solid lines). 
In contrast, Figure \ref{fig:den_wir} shows how the above changes with including the IR radiative force. Although the evolution before the transition (dashed lines) looks similar to that in Figure~\ref{fig:den_woir}, the late evolution (solid lines) is greatly modified; the collapse of the HII bubble is terminated at some point, so that the accreting gas gradually accumulates in the periphery of the bubble. The IR radiative force significantly modifies the density and thermal structures of accreting gas after the transition to quasi-steady phase.

\begin{figure*}
\begin{center}
\includegraphics[width=17cm,height=17cm]{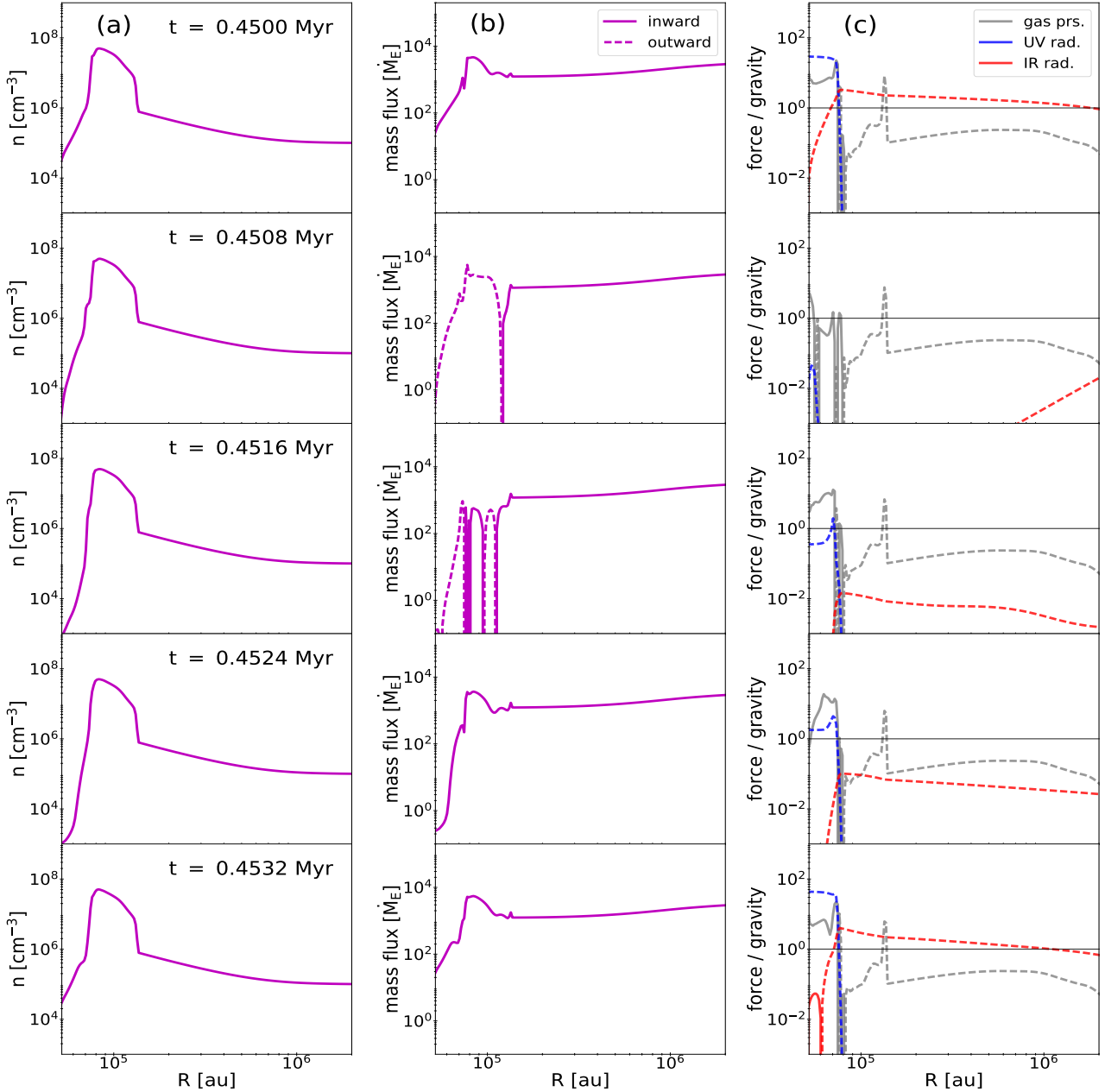}
\end{center}
\caption{Time variations of the accretion flow structure in the duration of $0.45 \lesssim t/\mr{Myr} \lesssim 0.453$ in model M5UVIR with $Z = 10^{-2} \ Z_\odot$ (see also Figure~\ref{fig:mdot}-b). 
The columns (a), (b), and (c) show the radial profiles of the number density, mass flux, and force balances, respectively. In each column, the different panels show the time-sequence every $8 \times 10^2$ years, which roughly corresponds to the free-fall time measured at the inner boundary.  
The solid and dashed lines in the columns (b) and (c) represent the inward and outward mass fluxes or forces.
In the column (b) the mass fluxes are normalized by the Eddington value $\medd$. 
In the column (c), the gray, blue, and red lines represent relative contributions of the gas pressure gradient, UV radiative force, and IR radiative force normalized by the BH gravity. 
}
\label{fig:Force}
\end{figure*}


To clarify the role of the IR radiative force in more detail, we further investigate the mass accretion after the transition into the quasi-steady state in M5UVIR model. 
As shown in Figure~\ref{fig:M5UVIR} (b), where the accretion rates are smeared over $10^4$ years, the accretion histories show only the small variability in the quasi-steady stage. 
However, Figure~\ref{fig:mdot} shows that the picture is totally changed once the accretion rates are averaged over the short duration $10^2$ years.
It is evident from Figures \ref{fig:mdot} (a) and (b) that the accretion rates actually very vigorously oscillate in the period of $\sim 10^3$ years.   


To understand what causes such a rapid oscillation of the accretion rate, we investigate variations of the accretion flow structure over one period of $\simeq 3000$ years after $t = 0.45~{\rm Myr}$, denoted with the thick black horizontal line segment in Figure \ref{fig:mdot} (b). 
The columns (a), (b), and (c) in Figure \ref{fig:Force} present such variations of radial profiles of the gas number density, mass flux, and force balances respectively. 
Starting at the epoch of $t = 0.45$ Myr, the density at the inner boundary is $\sim 10^5 \ \mr{cm^{-3}}$ and the BH rapidly accretes the gas with the mass influx of $\sim 10 \medd$. 
The resulting UV and IR radiative forces exceed the BH gravity everywhere, though the IR force almost balances with the gravity at the outer part as the dust temperature and IR opacity decline with increasing the radius. After that, the outflow is driven by the radiative forces in an inner part of $\lesssim 10^5 \ \mr{AU}$ by the epoch of $t = 0.4508$ Myr. 
The radiative forces disappear with the lack of the accretion, which allows for the gas to fall back again toward the central BH for $0.4508~{\rm Myr} < t < 0.4524~{\rm Myr}$. 
As the accretion rate rises, the radiative forces accordingly increase and eventually exceed the BH gravity by the epoch of $t = 0.4532$ Myr. 
The cycle is repeated thereafter
\footnote{During this cycle, although the density structure outside the HII region gradually evolves, as shown by the solid lines in Figure \ref{fig:den_wir}~(a), the behavior of gas accretion onto the central BH would not change significantly because it is determined almost only by the inner density structure.}.
Such a self-regulation process by the radiative forces clearly cause the oscillatory behavior of the accretion rates seen in Figure \ref{fig:mdot}.
%
%
It is worth noting that the period of the oscillation approximately corresponds to the free-fall time at the inner boundary; namely it would artificially be shorter with the smaller size of the sink region. However, we expect that our conclusion is not modified by this effect because the time-averaged accretion rate hardly changes with different positions of the inner boundary. See also Section \ref{sec:inner} for more details.
%
%


Although Figure \ref{fig:Force} shows that both the UV and IR radiative forces exceed the gravity, it is the IR force that predominantly regulates the mass accretion. In fact, such a regulation process disappears for cases M5UV, where only the UV radiative force is effective within the HII bubble. Previous studies show that the ram pressure of the accreting gas can be strong enough to overcome the repulsive UV radiative force \citep{Inayoshi16, Sakurai16}. 
The IR radiative force, in contrast, pushes back the accreting gas far outside of the HII bubble, where the gas has not obtained sufficient ram pressure yet.
It is thus the IR radiative force that regulates the dusty gas accretion onto the BH in the quasi-steady phase. 


Given the above argument, the mean accretion rate realized in the quasi-steady phase is expected to be governed by the balance between the IR radiative force and gravity. 
Such a balance is considered by invoking the Eddington luminosity defined 
with the dust IR opacity $\kappa_\mr{d,IR}$ instead of the standard Thomson scattering opacity, 
\begin{eqnarray}
\leddir &=& \frac{4 \pi G \mbh c}{\kappa_\mr{d,IR}} \nonumber \\
&=&  4.4 \times 10^7 \left ( \frac{\mbh}{10^3 \ \mr{M_\odot}} \right ) \left ( \frac{Z}{10^{-2} \ Z_\odot} \right )^{-1} \ L_\odot \ , \nonumber \\
\label{eq:leddir}
\end{eqnarray}
where we adopt $\kappa_\mr{d,IR} = 30 \ (Z/Z_\odot) \ \mr{cm^2 \ g^{-1}}$ supposing the dust temperature of $T_\mr{d} \sim 1000$~K, typical in the periphery of the HII bubble in our simulations. 
From Eqs. (\ref{eq:MtoL}) and (\ref{eq:leddir}), the corresponding Eddington mass accretion rate is described as
\begin{eqnarray}
\meddir =  
\begin{cases}
2~\medd \ \mr{exp} \left [ 0.67 \left ( \frac{Z}{10^{-2} \ Z_\odot} \right )^{-1} - 1 \ \right ] \\
\\
\ \ \ \ \ \ \ \ \ \ \ \ \ \  ({\rm for} \ Z < 6.7 \times 10^{-3} \ Z_\odot) \ , \\
\\
1.3~\medd \ \left ( \frac{Z}{10^{-2} \ Z_\odot} \right )^{-1} \ \ \ ({\rm otherwise}) \ .
\end{cases}
\label{eq:meddir}
\end{eqnarray}
In Figure \ref{fig:M5UVIR}(b), the red, green, and blue dotted lines represent $\meddir$ for the different metallicities of $Z = 10^{-3}, \ 10^{-2}, \ 10^{-1}~Z_\odot$, respectively. 
It is evident that the time-averaged mass accretion rates in the quasi-steady phase are in good agreement with $\meddir$. 


In summary, our results suggest that the condition $\meddir > \medd$ is necessary to realize the super-Eddington accretion via the quasi-steady accretion of dusty gas. 
This condition is rephrased as the critical gas metallicity of $\zcrit = 1.3 \times 10^{-2} Z_\odot$, above which the IR radiative force is too strong to realize the super-Eddington accretion, even in the quasi-steady accretion stage.

\subsection{Effects of Metal Cooling} \label{sec:mcool}

\begin{figure}
\begin{center}
\includegraphics[width=8cm,height=6cm]{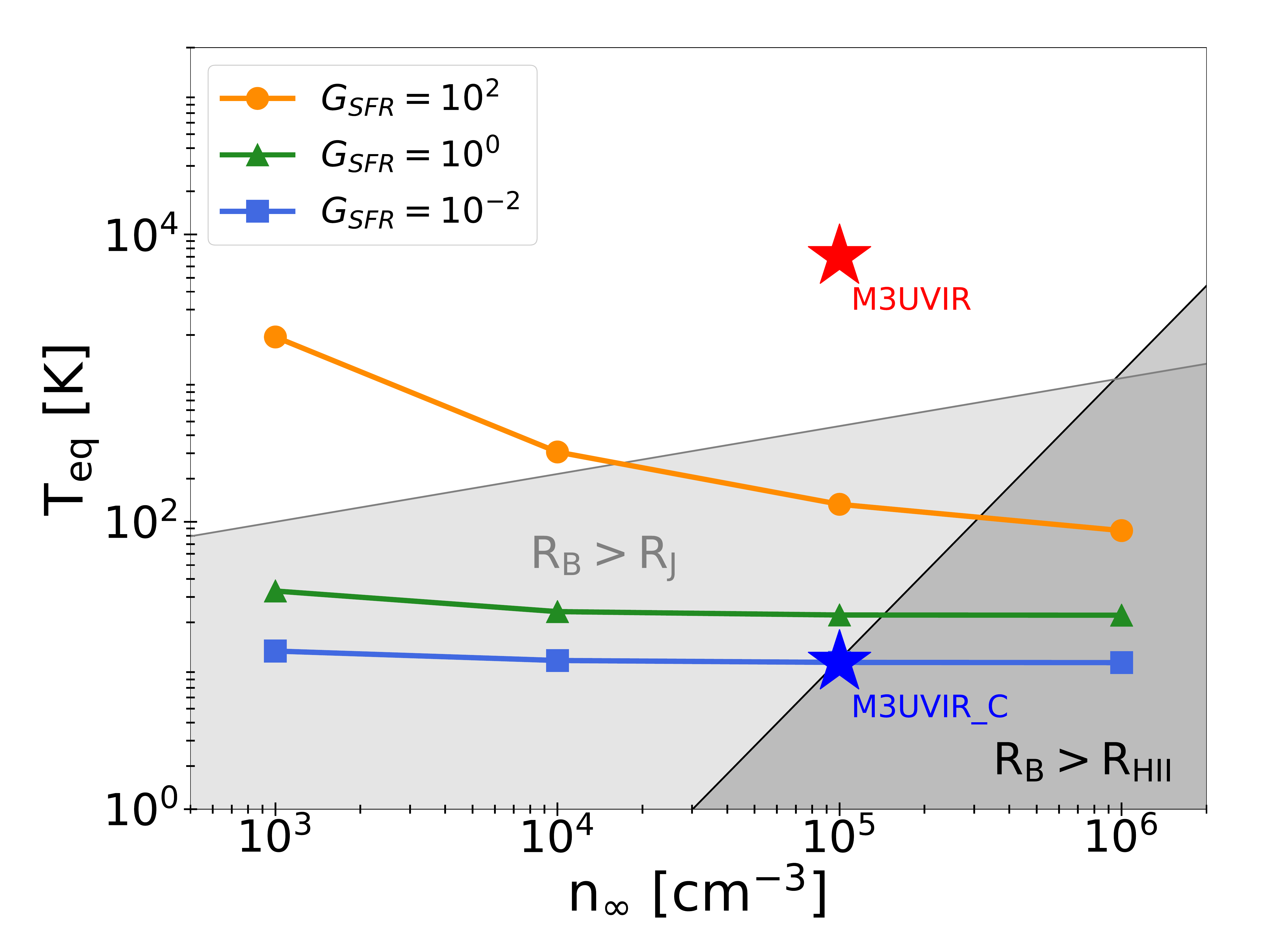}
\end{center}
\caption{The metallicity-dependent thermal equilibrium gas temperatures, and their resulting effects on the BH accretion process. 
The equilibrium temperatures at $Z = 10^{-2} \ Z_\odot$ are plotted as functions of the given ambient density $\ninf$ and surface density of the star formation rate $\gsfr$ (see text). 
The orange circles, green triangles, and blue squares represent the calculated equilibrium temperatures with the different star formation rates of $\gsfr = 10^{2}, \ 1, \ 10^{-2}$, which control the intensities of FUV and X-ray background fields. 
The red and blue star stars mark the ambient temperature and density used for M3UVIR and M3UVIR\_C models, respectively.  
The condition $\rb > \rhii$ for entering the quasi-steady accretion stage is satisfied in the dark-gray-shaded area with the BH mass $\mbh = 10^3 \ M_\odot$. Similarly, the condition $\rb > \rj$ is satisfied in the light-gray-shaded are, where the gas self-gravity will be effective in the BH accretion process (Eq. \ref{eq:condi4}).}
\label{fig:condition}
\end{figure}

We have assumed the constant ambient temperature $T_\infty = 10^4$~K above, supposing possible non-thermal turbulent motions in the ambient gas (Section~\ref{sec:initial}). 
However, the degree of turbulence in low-metallicity environments is yet to be studied. We hence consider limiting cases where the ambient temperature is just given by the thermal equilibrium. 
The equilibrium temperature $\teq$ depends on the metallicity owing to various heating and cooling processes via heavy elements contained in the gas and dust grains. 
We consider FUV and X-ray backgrounds from nearby star-forming regions as the heat sources. 
The intensities of these backgrounds are assumed to be proportional to $\gsfr$, the average surface density of the star formation rate normalized with the Galactic value $\Sigma_\mr{SFR} = 0.1 \rm M_\odot \ yr^{-1} \ kpc^{-2}$. The local intensities of the Galactic FUV and X-ray backgrounds are taken from \cite{Wolfire95}. 
Figure \ref{fig:condition} presents the gas equilibrium temperatures with $Z = 10^{-2} \ Z_\odot$ as functions of the ambient density $\ninf$ and star formation rate $\gsfr$. 
We see that, in the wide density range of $\ninf \sim 10^{3-6} \mr{cm^{-3}}$, the equilibrium temperature is $\sim 100$~K with active star formation, which is broadly expected in young gas-rich galaxies.

\begin{figure}
\begin{center}
\includegraphics[width=8cm,height=6cm]{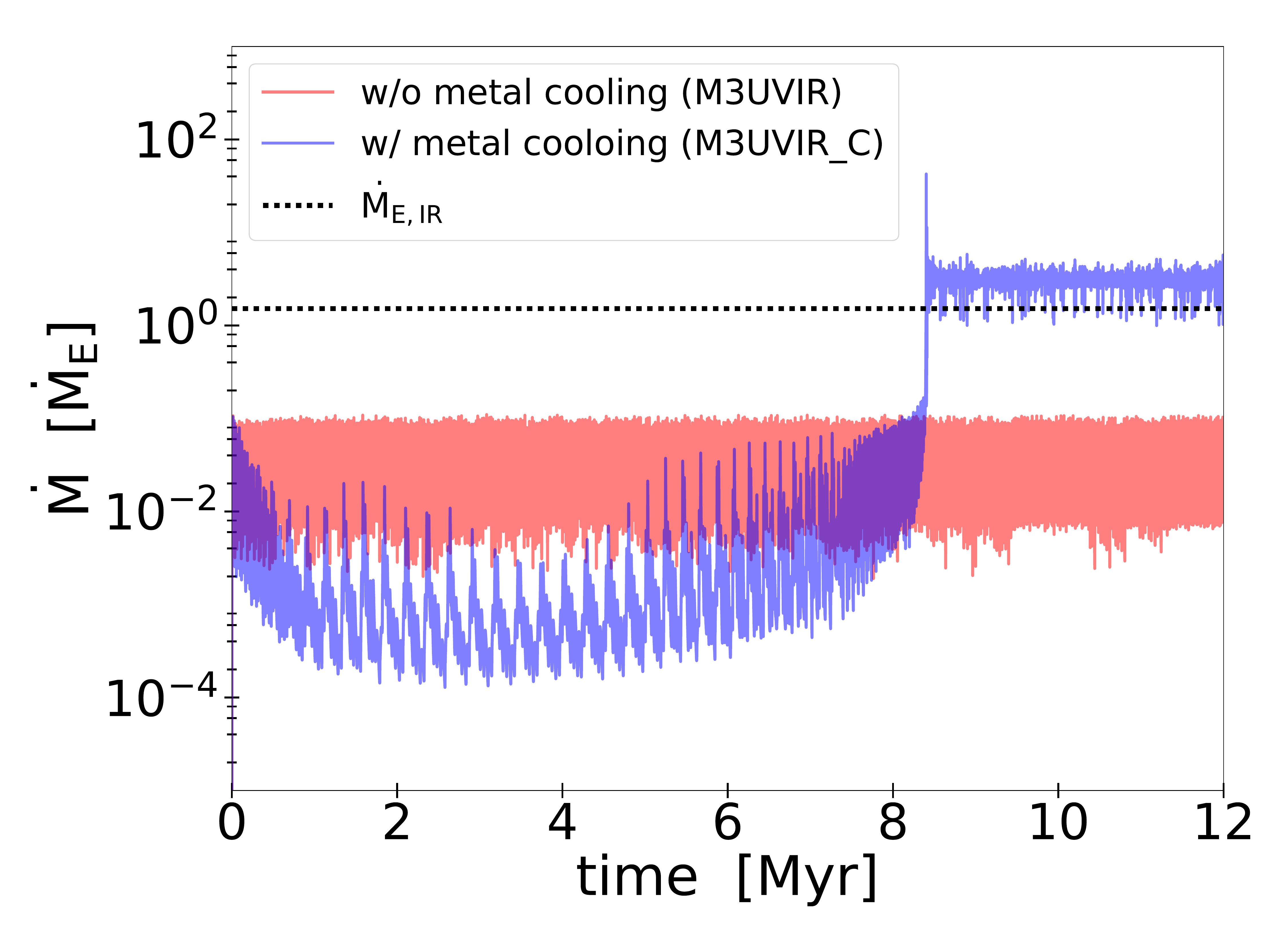}
\end{center}
\caption{Effects of using the thermal equilibrium temperature as the ambient temperature on the BH mass accretion histories. 
The red and blue lines represent M3UVIR and M3UVIR\_C models, respectively. 
The fixed ambient temperature $T_\infty = 10^4$~K is assumed for M3UVIR model, and the equilibrium temperature $\teq$ is used for M3UVIR\_C model. 
The time evolution of the mass accretion rates is presented in the unit of the Eddington value $\medd$ for the both cases.  The accretion rates shown in this figure are time-averaged for $10^2$ yrs. The black horizontal dotted line represents the Eddington accretion rate for IR radiative force given by Eq. (\ref{eq:meddir}).}
\label{fig:mcool}
\end{figure}

As expected from Eq.~(\ref{eq:condi1}), the quasi-steady accretion is more easily realized with the lower ambient temperature.  
We demonstrate this by comparing models M3UVIR and M3UVIR\_C (see Table 1), where the parameter sets are the same but for the different ambient temperatures: the fixed value $T_\infty = 10^4$~K for model M3UVIR, and the equilibrium temperature $\teq$ with the lowest star formation rate $\gsfr = 10^{-2}$ for M3UVIR\_C. In Figure \ref{fig:condition}, the red and blue star symbols represent these models. The condition $\rb > \rhii$ (Eq.~\ref{eq:condi1}) is satisfied in the dark-gray-shaded area, predicting the super-Eddington accretion is possible only for model M3UVIR\_C.
Figure \ref{fig:mcool} shows the calculated time evolution of the accretion rates in M3UVIR and M3UVIR\_C models, confirming the prediction given by Figure \ref{fig:condition}; 
 the transition to the quasi-steady accretion only occurs in model M3UVIR\_C.
The mean accretion rate after the transition well converges to the Eddington value given by Eq. (\ref{eq:meddir}), $\meddir$.


%
%
%
%

\section{Dusty Super-Eddington Accretion in Young Galaxies} \label{sec:gal}

Our results have suggested the following two conditions are required for the super-Eddington accretion to occur with the dusty gas: 
(i) $\rb > \rhii$ (Eqs. \ref{eq:condi1} and \ref{eq:condi2}), with which the quasi-steady accretion is possible against the thermal pressure and UV radiative forces within the HII bubble, and 
(ii) $Z \lesssim 10^{-2} \ Z_\odot$, with which the accretion rate in the quasi-steady stage $\meddir$ exceeds the normal Eddington value $\medd$.  
We here investigate whether the two conditions are actually satisfied in young galaxies.


First, we consider the condition of $Z \lesssim 10^{-2} \ Z_\odot$. 
Such metal-poor environments will be preferentially found in the early universe. 
Unfortunately, observations on the chemical evolution of galaxies are still limited for $z \lesssim 4$ \citep[e.g.,][]{Mannucci10, Troncoso14, Hunt16, Onodera16}.
Recent high-resolution simulations have been revealing the very early epoch of the chemical enrichment in the first galaxies \citep[e.g.,][]{Wise12, Ricotti16}, though their results are not suitable for our statistical assessment owing to their small simulation boxes.
We then rely on recent cosmological simulations dedicated to the early galaxy formation by \cite{Sarmento18}, who use a large simulation box with over 10 Mpc $h^{-1}$ on a side. Their results explain statistical properties of young galaxies, such as the rest-frame UV luminosity functions observed at $z \geq 7$.
They predict the relation between the metallicity in a galaxy and the virial mass of its host halo $\mvir$ at $z$ = 7-15, which is approximately fitted with
\begin{eqnarray}
\mr{log}(Z/Z_\odot) = - 0.5 - 0.085 \ z + 0.48 \ m - 0.13 \ m^2  + 0.058 \ m^3 \ , \nonumber \\
\label{eq:mvzr}
\end{eqnarray}
where $m \equiv \mr{log}(\mvir/10^{10}~M_\odot)$ and the metallicity $Z$ represents the mean value averaged over a galaxy. Although the chemical enrichment proceeds more rapidly at the center than in an outer part, we only consider BHs accreting the gas with the mean metallicity $Z$. Note that, in this paper, we consider not only BHs residing at the galactic centers, but also those free-floating in galactic gas discs (also see discussions in Section~\ref{sec:angular}).
We argue that the accretion onto such free-floating BHs is rather the key to understand the quick growth of the seed BHs in the early universe \citep[also see Section 5.1 in][]{Sugimura18}.

\begin{figure}
\begin{center}
\includegraphics[width=8cm,height=6cm]{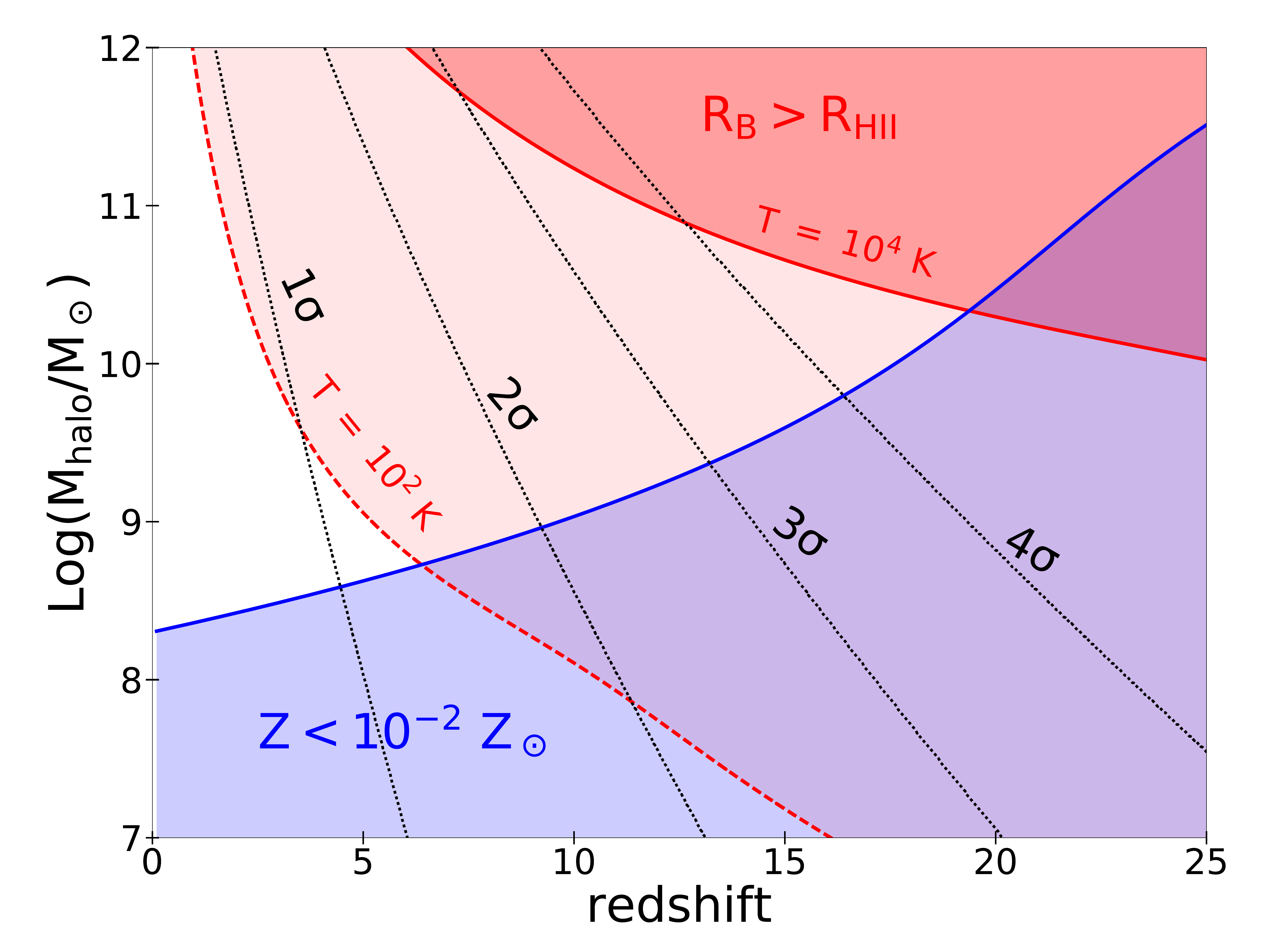}
\end{center}
\caption{Young galaxies as potential sites of the dusty super-Eddington accretion, in the parameter space of redshifts and host halo masses. The blue-shaded region represents the low-metallicity galaxies with $Z \lesssim 10^{-2} \ Z_\odot$, for which the condition $\meddir \gtrsim \medd$ is satisfied according to Eq. (\ref{eq:mvzr}). 
The red solid and dashed lines are the borders on which the Str\"omgren radius $\rhii$ is equal to the Bondi radius $\rb$ with different ambient temperatures $T = 10^4$~K and $10^2$~K, where a particular BH mass $\mbh = 10^3 \ M_\odot$ is assumed. 
The condition $\rb > \rhii$ is met in the red-shaded regions delineated by these lines.
The areas where the red and blue regions are overlapped denote the potential sites of the dusty super-Eddington accretion, as the above two conditions are both satisfied.  
The thin dotted lines represent the virialization redshifts for different halo masses, where the variance of original density fluctuations correspond to 1$\sigma$, 2$\sigma$, 3$\sigma$, and 4$\sigma$.}
\label{fig:mhalo-z}
\end{figure}

Figure \ref{fig:mhalo-z} shows that, with Eq. (\ref{eq:mvzr}), galaxies with $Z \lesssim 10^{-2}~Z_\odot$ are expected to be hosted by dark halos located in the blue-shaded area. 
At the epoch of $z \simeq 10$, for instance, such low-metallicity galaxies are in halos less massive than $\sim 10^9 \ M_\odot$. 
Figure \ref{fig:mhalo-z} also shows that these halos at $z \simeq 10$ come from the density fluctuations with the mass variance of $\lesssim 2~\sigma$. Note here that Eq. (\ref{eq:mvzr}) might provide a poor prediction for the chemical evolution in halos with $\mvir \gtrsim 10^{12}~M_\odot$ because of the scarcity of such massive halos. This does not affect our argument on young galaxies hosted with the less massive halos, as shown in Figure \ref{fig:mhalo-z}. 


Next, we investigate the condition $\rb > \rhii$, which is equivalent to Eq. (\ref{eq:condi1}) for low-metallicity galaxies with $Z \lesssim 10^{-2} \ Z_\odot$ according to Eq. (\ref{eq:zdcrit}).
We consider a BH in a purely gaseous disk, which is generally valid in early universe, and approximate the disk radius $\rd$ with the virial radius of the host halo $\rvir$ as
\begin{eqnarray}
\rd &=& \lambda \ \rvir \nonumber \\
&=& \lambda \left ( \frac{2}{\Delta_\mr{vir}(z)} \frac{G \mvir}{H(z)^2} \right )^{1/3} \nonumber \\
&\simeq& 5.1 \times 10^{7} \ \mr{AU} \mvten^{1/3} \zten^{-1} \ ,
\label{eq:rd}
\end{eqnarray}
where we have adopted the spin parameter of the host halo $\lambda = 0.035$, which is the median value obtained from cosmological N-body simulations by \cite{Bullock01}, and the standard cosmological parameters \citep{Spergel07}.
We here assume the mass of the gas disk as $\mg = 0.1 \mvir$ following the standard baryon to dark matter mass ratio $\mr{f}_B \simeq 0.17$. 
Further considering the hydrostatic balance in the vertical direction, the disk scale height is written as 
\begin{eqnarray}
\zd &=& \frac{c_\mr{s}^2}{2\pi G \sgmd} \nonumber \\
&\simeq& 1.5 \times 10^{5} \ \mr{AU} \mvten^{-1/3} \zten^{-2} \left ( \frac{T}{10^4~K} \right ) \ ,
\label{eq:zd}
\end{eqnarray}
where $\sgmd = \mg/\pi\rd^2$ is the surface mass density and $c_\mr{s}$ is the sound speed.
The gas number density within the disk is estimated as
\begin{eqnarray}
\nd &=& \frac{\mg}{2 \pi m_\mr{p} \zd \rd^2} \nonumber \\
&\simeq& 1.5 \times 10^5 \ \mr{cm^{-3}} \mvten^{2/3}  \tfour^{-1} \zten^4 \ . \nonumber \\
\label{eq:nd}
\end{eqnarray}
Finally, substituting Eq. (\ref{eq:nd}) into $\ninf$ in Eq. (\ref{eq:condi1}), 
the condition $\rb > \rhii$ is rewritten as
\begin{eqnarray}
& & \left ( \frac{\mbh}{10^5 \ M_\odot} \right ) \mvten^{2/3} \zten^{4} \gtrsim \ \ \ \ \ \ \ \ \ \ \ \ \ \ \ \  \nonumber \\
& & 0.67 \ \left ( \frac{T}{10^4 \ \mr{K}} \right )^{3/2} \left(1 + 7.1 \times \frac{Z}{10^{-2} \ Z_\odot} \right)^{-1/2} \ , 
\label{eq:condi3}
\end{eqnarray}
where we have assumed $\thii = 7 \times 10^4 \ \mr{K}$ to drop the $\thii$-dependence. We evaluate this condition with using a fixed ambient temperature $T = 10^4$~K as adopted for many cases in our simulations (Table 1). In Figure \ref{fig:mhalo-z}, the thick red-shaded region represents the halos which satisfy the condition $\rb > \rhii$ for such cases with $\mbh = 10^3 \ M_\odot$. We find from the figure that the quasi-steady accretion is only possible for massive halos with $M_{\rm vir} \gtrsim 10^{11}~M_\odot$ at $z \gtrsim 12$. These halos correspond to rare density fluctuations with the mass variance larger than $3\sigma$.


The area where the thick red- and blue-shaded regions overlap denotes the galaxies where the dusty super-Eddington accretion onto BHs is realized, satisfying both the conditions. 
We see that the area corresponds to very limited cases with $\mvir > 10^{10} \ M_\odot$ and $z \gtrsim 20$, rarer than $4 \sigma$ density fluctuations. 
This is problematic because the number density of these halos is much lower than that of SMBHs observed at $z \simeq 6$, $\sim 1~h^3~{\rm Gpc}^{-3}$ \citep[e.g.,][]{Reed07}.


However, the above estimate sensitively depends on our choice of the ambient temperature.
For instance, consider cases with $T = 100$~K as investigated in Section~\ref{sec:mcool}. 
We note that it is comparable to the typical kinetic temperature of circum-nuclear disks in nearby galaxies \citep[e.g.,][]{Davies12, Izumi13, Viti14}.
In Figure \ref{fig:mhalo-z}, the thin red-shaded region presents the resulting parameter space allowing $\rb > \rhii$. Now the red-shaded region has the much larger area overlapping with the blue-shaded region than before. The conditions of $\rb > \rhii$ and $Z < 10^{-2}~Z_\odot$ are both met for galaxies even in $\sim 2 \sigma$-$3 \sigma$ halos at $z \gtrsim 10$.
For instance, a galaxy hosted by a $3 \sigma$ halo at $z \simeq 15$ is expected to have $Z \sim 10^{-3}~Z_\odot$, and to realize very rapid BH accretion with $\dot{M} \sim \meddir \sim 10^3 \medd$. 
The number density of such halos is more than 100 times larger than that of SMBHs found at $z \simeq 6$. 



\section{DISCUSSIONS} \label{sec:discussion}

\subsection{Dust Sublimation Radius and Inner Boudary} \label{sec:inner}

We here discuss effects of the dust sublimation front, which is not modeled in our simulations. For the quasi-steady accretion stage, in which the BH accretion disk radiates at $\leddir$, the dust sublimation radius is evaluated as
\begin{eqnarray}
\rsb &=& \left ( \frac{\leddir}{4 \pi \sigma_\mr{SB} T^4_\mr{d}} \right )^{1/2} \nonumber \\
&\sim& 10^4 \left ( \frac{\mbh}{10^5 \ \mr{M_\odot}} \right )^{1/2} \left ( \frac{Z}{10^{-2} \ Z_\odot} \right )^{-1/2} \left ( \frac{T_\mr{sb}}{10^{3} \ \mr{K}} \right )^{-2} \ ~\mr{AU} \ , \nonumber \\
\label{eq:rsb}
\end{eqnarray}
where $\sigma_\mr{SB}$ is Stefan-Boltzmann constant and $T_\mr{sb}$ is the dust sublimation temperature for graphite and silicate grains. The above is somewhat smaller than our fiducial choice of the inner boundary $R_\mr{in} = 1/30~\rb \sim 5 \times 10^4 \mr{AU}$, so that it is not resolved in our simulations. We have reduced the computational costs in return for such a limitation. 


\cite{Krumholz18} demonstrates that resolving the dust sublimation front is a key to correctly follow the dusty gas accretion under the radiative feedback from a bright source. It is reasonable because poorly resolving the front results in relatively underestimating the ram pressure of the accretion flow in comparison to the outward radiative force (also see Sec.~\ref{sec:rad_ir}). As a result, one may overestimate the strength of the radiative feedback for such a case. In order to examine this effect, we preform a test simulation with the same setting as in M5UVIR model, but for the smaller inner boundary $R_\mr{in} = 10^4 \mr{AU}$.
The test calculation confirms that the quasi-steady accretion occurs at the rate of $\meddir$ as in the fiducial case. Only a difference is that the HII region and surrounding shell-like structure is confined into a smaller region, within $\simeq 3 \times 10^4$~AU from the center. 

Note that the requirement proposed by \cite{Krumholz18} mainly supposes cases where the UV radiative force is strong enough to halt the accretion. Resolving the dust sublimation front is critical for such a case because most of UV photons are absorbed almost at once there to create a large impulsive outward force. For our cases, however, it is the IR force that regulates the accretion rates in the quasi-steady state. Recall that the UV force is not strong enough to prevent the quasi-steady accretion in M4UV and M5UV models, where its effect is even overestimated owing to our limited spatial resolution.


\subsection{Multi-Dimensional Effects} \label{sec:multi}

While our 1D simulations require $Z < 10^{-2} \ Z_\odot$ for dusty super-Eddington accretion, it may be too stringent in light of recent observations. 
For instance, it has been suggested that the SMBHs found at $z \gtrsim 6$ accrete the gas nearly at the Eddington rates \citep{Mortlock11, Banados18}. 
With estimated metallicities of their host galaxies $Z \sim 0.1 \ Z_\odot$ \citep[e.g.,][]{Venemans12, Venemans17}, it corresponds to $\dot{M} \sim 10 \meddir$.
We here discuss several effects that are not incorporated in our simulations and may allow $\dot{M} > \meddir$.

\subsubsection{Self-gravity of gas} \label{sec:self-grav}

First, we estimate the effect of self-gravity of the gas by comparing the Bondi radius to the Jeans length defined as
\begin{eqnarray}
\rj &=& \sqrt{\frac{\pi}{G\rho}}c_\mr{s} \nonumber \\
&=& 1.1 \times 10^6 \left ( \frac{\tinf}{10^4 \ \mr{K}} \right )^{1/2} \left ( \frac{\ninf}{10^5 \ \mr{cm^{-3}}} \right )^{-1/2} \ \mr{AU} \ . \nonumber \\
\label{eq:rj}
\end{eqnarray}
For cases with $\rj < \rb$, the gas within the Bondi radius fragments into clumps owing to the gravitational instability. 
The condition $\rj < \rb$ is rewritten using Eqs. (\ref{eq:rbondi}) and (\ref{eq:rj}) as
\begin{eqnarray}
\left ( \frac{\tinf}{10^4 \ \mr{K}} \right ) \lesssim \left ( \frac{\mbh}{10^5 \ \mr{M_\odot}} \right )^{2/3} \left ( \frac{\ninf}{10^5 \ \mr{cm^{-3}}} \right )^{1/3}  \ . \nonumber \\
\label{eq:condi4}
\end{eqnarray}
In Figure \ref{fig:condition}, the light gray-shaded region denotes the parameter space of the ambient temperature and density where $\rj < \rb$ is satisfied with $\mbh = 10^3 \ M_\odot$. 
The figure shows that the gas self-gravity is always effective when the ambient temperature is given by the thermal equilibrium values, with the temperatures lower than $\sim 100 - 10^3$~K moderately depending on the density. This contains cases where the transition to the quasi-steady accretion stage is expected with $\rb > \rhii$. In such cases, the accreting gas actually has clumpy structure created by the gravitational fragmentation. 
The diffuse IR radiation are expected to preferentially escape through only sparse regions among the clumps \cite[e.g.,][]{Jumper18}. The resulting IR radiative force might not efficiently halt the accretion flow toward the central BH, and allow $\dot{M} > \meddir$.

%
%
Of course, this is just an aspect of possible effects caused by the gas self-gravity. If a part of the clumps is turned into stars, it may take quite a long time for them to reach a BH owing to the so-called loss-cone filling. Furthermore, various stellar feedback effects such as supernova explosions would significantly modify the physical states of the gas, which might even reduce the accretion rates onto a BH (also see Sec.~\ref{sec:angular} below). 
%
%


\subsubsection{Angular momentum of accreting gas} \label{sec:angular}

We have focused on cases where the size of a BH accretion disk is much smaller than the Bondi radius, and ignored angular momentum in our simulations. \cite{Sugimura18} show that, with 2D RHD simulations, such an approximation is valid if the disk is $\lesssim 0.01$ times smaller than the Bondi radius. They estimate that this is the case for free-floating BHs accreting the clumpy ISM, where the net gain of the angular momentum is due to the density and velocity fluctuations over the spatial scale of the Bondi radius \citep[e.g.,][]{Ipser77, Ioka17, Matsumoto18}. On the other hand, BHs located at galactic centers generally accrete the gas with more amounts of the angular momentum. \cite{Sugimura18} also show that, for such a case, accretion rates are reduced by ${\cal O}(\alpha) \sim 0.01-0.1$ with the $\alpha$-viscosity parameter. The gas angular momentum is thus a potential obstacle for efficient BH mass growth via the super-Eddington accretion.


However, the angular momentum may also play the opposite role of promoting the accretion with the dusty gas. With a geometrically thin accretion disk forming with the angular momentum, the diffuse IR photons preferentially escape in the vertical direction of the disk. The accretion flow within the disk is hardly disturbed by the IR radiative force. 
This, known as ``flashlight'' effect \citep[e.g.,][]{YB99, Kuiper10}, mitigates the significance of the IR radiative force, which in 1D is strong enough to limit the accretion rate as $\dot{M} \sim \meddir$. Unfortunately, it is still uncertain how the net effects of the angular momentum modify the accretion rate.


In order to fully understand the role of the angular momentum, it is necessary to incorporate physical processes that cause the torque, instead of relying on the phenomenological $\alpha$-viscosity prescription. We expect that, for cases of the dusty BH accretion, some parts of the accretion disk would be gravitationally unstable owing to efficient radiative cooling via heavy elements. 
Non-axisymmetric spiral structure develops to cause the gravitational torque for such a case \citep[e.g.,][]{Kuiper11, Hosokawa16}. An unstable disk may fragment into clumps, which further collapse into stars. Various feedback effects from the new-born stars heat the gas, and finally regulate the stability of the disk. Analytic models predict that rapid accretion onto BHs is realized through such a complex interplay \citep[e.g.,][]{Thompson05, Kawakatu08}, which is to be verified in future 3D simulations.

\subsubsection{Supersonic turbulence in ambient gas} \label{sec:turb}

Finally, we consider effects of supersonic turbulence in the ambient gas.  Modern simulations show that such strong turbulence ubiquitously occurs in young galaxies \citep[e.g.,][]{Wise07, Hopkins11}. It is easily driven by the gravitational energy release via the cosmological mass assembly, and also by various feedback from star formation activities. As demonstrated in Figure \ref{fig:mhalo-z}, the supersonic turbulence can limit the transition to quasi-steady accretion by effectively increasing the sound speed. The turbulence in the accreting gas may prevent the efficient BH mass growth. 


However, it is to be checked that the turbulence is approximated using the effective sound speed, because it generally acts as an anisotropic pressure. Recent studies further suggest that rapid accretion onto BHs is actually possible when the turbulence dissipates \citep[e.g.,][]{Hobbs11, Faber18}.
According to their 3D simulations, converging turbulent flows can produce dense streams and clumps in the ambient medium. Further collisions of these overdense structures induce their orbital cascades, leading to efficient mass supply for a BH. Moreover, density fluctuations created by turbulent motions are expected to weaken the effect of IR radiative force against the accretion flow (see also Section \ref{sec:self-grav}). 
It is a task for future studies to clarify roles of the turbulence in the BH mass growth. Understanding nature of the low-metallicity ISM, which is still limited, is indispensable for that purpose \citep[e.g.,][]{Inoue15}.


\section{SUMMARY AND CONCLUSION} \label{sec:summary}

In this paper, we have investigated how the rapid mass growth of seed BHs may be possible via the super-Eddington accretion in the early universe. In particular, we have focused on effects of dust grains contained in the accreting gas. 
Since the dust grains have larger opacity than the gas, dynamics of the accreting gas is substantially affected by enhanced radiative feedback from accreting BHs. 
We have performed a suite of 1D RHD simulations to investigate such effects under the assumption of spherical symmetry. Previous studies show that, with the primordial gas composition, the super-Eddington quasi-steady accretion occurs when the Bondi radius is larger than the typical size of an HII bubble around the central BH, $\rb > \rhii$. We investigate such cases with various metallicities, and show that $\rb > \rhii$ is not the sufficient condition for the super-Eddington accretion with the dusty ambient medium.


In order to isolate effects of the radiative forces described above, we have first fixed the ambient temperature at $T_\infty = 10^4$~K, regardless of different metallicities. 
This may approximate the ambient gas filled with supersonic turbulence.
Our results show that the radiative force caused by the IR dust thermal emission plays a major role to regulate mass accretion. The mean accretion rate onto a BH is lower with higher metallicity owing to its effect. For instance, for a $10^5~M_\odot$ BH embedded in the ambient gas with the density $10^5~{\rm cm}^{-3}$, the accretion rate with $Z = 0.1~Z_\odot$ is $\sim 5$ orders of magnitude smaller than that with $Z=0$ (Fig.~2). 
In the accretion flow outside of the HII bubble, the BH gravity is nearly balanced with the radiative force via the IR dust thermal emission.  
The luminosity of the accretion disk is thus regulated at the Eddington value, which is defined with dust IR opacity instead of the Thomson scattering opacity (Eq.~\ref{eq:leddir}). 
The resulting accretion rate is also given by the corresponding Eddington value $\dot{M}_{\rm E,IR}$ (Eq.~\ref{eq:meddir}). With these results, we have derived the critical metallicity $Z \sim 10^{-2} \ Z_\odot$, above which $\dot{M}_{\rm E,IR}$ is lower than the normal Eddington accretion rate. In addition to $\rb > \rhii$, satisfying $Z \leq 10^{-2} \ Z_\odot$ is also necessary for the super-Eddington accretion with the dusty gas.


Next, we have considered cases where the ambient temperatures are no longer fixed, but given by the thermal equilibrium values $T_{\rm eq}$. 
With $Z = 10^{-2} \ Z_\odot$, for instance, the equilibrium temperature is typically $\sim 100$~K in the dense medium with $\ninf \gtrsim 10^{3} \mr{cm^{-3}}$. 
Since the Bondi radius is larger with the lower temperature, the condition $\rb > \rhii$ is more easily met than with $T_\infty = 10^4$~K. From the viewpoint of this effect, finite amounts of heavy elements in the ambient gas rather promote the rapid accretion growth of the seed BHs. In summary, the accreting gas with the metallicity slightly lower than the critical value $10^{-2}~Z_\odot$ may provide the most optimal condition for the super-Eddington accretion. 


We have further discussed which young galaxies may provide such favorable conditions with $\mbh = 10^3 \ M_\odot$ seed BHs.
Specifically, by using a simple galaxy formation model, we have assessed the two conditions for super-Eddington accretion (i) $\rb > \rhii$ and (ii) $Z < 10^{-2} \ Z_\odot$, as a function of redshift and virial mass of halos. With the fixed ambient temperature $T_\infty = 10^4$~K, we have found that the dusty super-Eddington accretion is only possible with galaxies hosted by very rare halos corresponding to 4$\sigma$ density fluctuations. The number density of these halos is far below that of SMBHs observed at $z \simeq 6$, $\sim 1~h^3~{\rm Gpc}^{-3}$. With limiting cases with $T_\infty = 100~{\rm K} \sim T_{\rm eq}$, however, the possible parameter space dramatically extends. In this case, galaxies even in $\sim 2 \sigma - 3 \sigma$ halos at $z \gtrsim 10$ are potential sites of the super-Eddington accretion. For instance, the rapid BH mass growth with $\dot{M} \sim 10^3 \ \medd$ is possible in a galaxy hosted by a 3$\sigma$ halo at $z \simeq 15$. The number density of such halos is much larger than $1~h^3~{\rm Gpc}^{-3}$. Therefore, the super-Eddington accretion with the cold ($T \sim 100$~K) dusty gas is a possible pathway to quickly form SMBHs in the early universe.


The above assessment clearly suggests that understanding the exact nature of the low-metallicity interstellar medium, which is probably dominated by self-gravitating turbulence, is essential to explore the possibility of the super-Eddington BH growth with dusty gas.  
The resulting clumpiness of the accreting gas is also a key to evaluate the strength of the radiative feedback effects. Although such effects are to be investigated in future studies, we believe that our current work provides a first step toward such directions. 

\section*{Acknowledgements}
We thank Hajime Fukushima for providing his code to calculate the virialization redshifts of halos associating with different overdensities shown in Figure \ref{fig:mhalo-z}. We are also grateful for fruitful discussions with Masayuki Umemura, Ken Ohsuga, Hidenobu Yajima, Nozomu Kawakatu, Kohei Inayoshi, and Kazu Omukai. This work is supported in part by MEXT Grant-in-Aid for Scientific Research (No. 16H05996 and 17H06360 for TH). RK acknowledges financial support via the Emmy Noether Research Group on Accretion Flows and Feedback in Realistic Models of Massive Star Formation funded by the German Research Foundation (DFG) under grant no. KU 2849/3-1.

\bsp	
\label{lastpage}
\end{document}